\documentclass[aip,pop,amsmath,reprint,floatfix]{revtex4-1}
\usepackage{dcolumn}
\usepackage{bm}
\usepackage{graphicx}
\usepackage{hyperref}

\begin{document}

\title{Convective Raman Amplification of Light Pulses Causing Kinetic Inflation in Inertial Fusion Plasmas} 



\author{I. N. Ellis}
\email{ellis@physics.ucla.edu}
\affiliation{Lawrence Livermore National Laboratory, Livermore, California 94550, USA}
\affiliation{University of California, Los Angeles, California 90095, USA}

\author{D. J. Strozzi}
\affiliation{Lawrence Livermore National Laboratory, Livermore, California 94550, USA}

\author{B. J. Winjum}
\author{F. S. Tsung}
\affiliation{University of California, Los Angeles, California 90095, USA}

\author{T. Grismayer}
\affiliation{University of California, Los Angeles, California 90095, USA}
\affiliation{Grupo de Lasers e Plasmas, Instituto Superior T\'{e}cnico, 1049-001 Lisboa, Portugal}

\author{W. B. Mori}
\author{J. E. Fahlen}
\affiliation{University of California, Los Angeles, California 90095, USA}

\author{E. A. Williams}
\affiliation{Lawrence Livermore National Laboratory, Livermore, California 94550, USA}


\date{\today}

\begin{abstract}
We perform 1D particle-in-cell (PIC) simulations using OSIRIS, which model a short-duration ($\sim500\omega_0^{-1}$ FWHM) scattered light seed pulse in the presence of a constant counter-propagating pump laser with an intensity far below the absolute instability threshold.  The seed undergoes linear convective Raman amplification and dominates over fluctuations due to particle discreteness.  Our simulation results are in good agreement with results from a coupled-mode solver when we take into account special relativity and the use of finite size PIC simulation particles.  We present linear gain spectra including both effects.  Extending the PIC simulations past when the seed exits the simulation domain reveals bursts of large-amplitude scattering in many cases, which does not occur in simulations without the seed pulse.  These bursts can have amplitudes several times greater than the amplified seed pulse, and we demonstrate that this large-amplitude scattering is the result of kinetic inflation by examining trapped particle orbits.  This large-amplitude scattering is caused by the seed modifying the distribution function earlier in the simulation.  We perform some simulations with longer duration seeds, which lead to parts of the seeds undergoing kinetic inflation and reaching amplitudes several times more than the steady-state linear theory results.  Simulations with continuous seeds demonstrate that the onset of inflation depends on seed wavelength and incident intensity, and we observe oscillations in the reflectivity at a frequency equal to the difference between the seed frequency and the frequency at which the inflationary SRS grows.

\end{abstract}



\maketitle 

\section{Introduction}
Backward stimulated Raman scattering\cite{drake:parametric, forslund:theory, *forslund:simulation} (BSRS) in plasmas, in which an incident light wave in a plasma decays into a backward-propagating light wave and a forward-propagating plasma wave, has been a subject of much study, in large part because it scatters light away from the target in inertial confinement fusion (ICF).\cite{atzeni:physics, lindl:physics}  Early research focused on relatively high intensities, where growth was in the weakly damped convective and absolutely unstable regimes, and saturation occurred due to wave-breaking and/or pump depletion.\cite{koch:nonlinear, estabrook:nonlinear}  The competition between back-, forward-, and side-scatter was also investigated.\cite{Estabrook:Theory, kruer:raman}  As laser and plasma parameters for ICF evolved, research in BSRS shifted to the strongly damped regime.  In modern experiments, SRS typically occurs at densities and temperatures for which $k\lambda_{De} \gtrsim 0.3$, where Landau damping is significant.  In this kinetic regime, the measured BSRS reflectivities can greatly exceed the values from linear theory calculations, e.g. in the single-hot-spot experiments of Ref. \onlinecite{montgomery:recent}.  3D paraxial-envelope simulations with linear damping, however, correctly modeled the intensity threshold for SRS in experiments with a smoothed, multi-speckle beam in a pre-formed uniform hohlraum plasma.\cite{froula:observation}  A process called kinetic inflation was proposed to explain the single-hot-spot results.\cite{vu:transient,*vu:kinetic,vu:inflation,strozzi:kinetic} In kinetic inflation, a small-amplitude plasma wave excited in the strongly damped convectively unstable regime can trap electrons, modifying the distribution function so that the kinetic damping of the plasma wave is greatly reduced or vanishes.\cite{oneil:collisionless, morales:nonlinear}  Therefore, for the same incident laser intensity, SRS can then transition to the weakly damped or absolutely unstable regime.\cite{winjum:dissertation}  There has also been recent work on how BSRS in the kinetic regime can saturate due to nonlinear frequency shifts\cite{vu:transient,*vu:kinetic,Winjum:Effects} or related trapped-particle instabilities\cite{brunner:trapped} caused by electron trapping.  Recent research has demonstrated the importance of the propagation and evolution of plasma wave packets, including how the reflected light can occur in bursts spaced proportionally to the inverse of the nonlinear frequency shift.\cite{Winjum:Effects}  The latest research has also demonstrated that hot electrons and back- and side-scattered SRS produced by one speckle interact with neighboring speckles, causing the speckles to self-organize and produce coherent bursts of SRS.\cite{Yin:Self-Organized}  Until recently, little work has explored the possibility of scattered light, plasma waves, or the resulting changes to the electron distribution in one region of space or time enhancing SRS at different times or locations.

In this paper, we make a detailed comparison of the linear amplification of a well defined counter-propagating seed pulse using coupled-mode theory and OSIRIS PIC\cite{fonseca:osiris} simulations.  We then explore how this seed pulse can trigger large reflectivities after it has left the plasma using OSIRIS simulations. We consider situations in which no BSRS occurs with only the pump (no seed is used). The seed intensity and pulse length are varied. For short seed pulses ($\sim500\omega_0^{-1}$ FWHM), the seed pulse is linearly amplified as it transits the box.  This amplification agrees with linear theory when appropriately modified to take into account special relativity and the use of finite-size particles in PIC codes,\cite{dawson:particle,birdsall:plasma} such as OSIRIS. We simulate ICF-relevant laser and plasma conditions, and demonstrate that special relativity increases the linear gain and shifts down the scattered light wavelength.  We find, for these short seed pulses, that kinetic inflation occurs after the seed pulse leaves the box.  The timing and amplitude of the first peak in reflectivity after the seed pulse depends on the duration and intensity of the seed pulse.  We examine the trapped particles to verify that kinetic inflation is occurring and that the bounce period is consistent with the Langmuir wave (which we also call the plasma wave) amplitude.

For longer seed pulses, the inflationary burst of scattered light overlaps with the seed.  Under these conditions, the measured gain of the seed can reach several times the steady-state linear gain value when the seed wavelength is near the peak of the gain spectrum. We also examine the onset of inflationary scattering and the bursty nature of BSRS using continuous seeds.  Non-resonant seeds, which are not at the peak of the linear gain curve, require higher incident intensity to cause inflation.  We also observe that, when the seed frequency is non-resonant, the reflected light is modulated with a period inversely proportional to the difference between the seed and resonant frequencies.  (In this paper, we use the term ``resonance," where $1+\chi_r=0$ for electrostatic waves, and ``peak gain" interchangeably.) After inflation sets in, the measured gain of the seed \textit{decreases} with incident seed intensity due to pump depletion.

The paper is outlined as follows. We present the simulation geometry and plasma conditions in Section \ref{geom}, and discuss the linear theory of convective BSRS gain and its relativistic and PIC modifications in Section \ref{gain}.  We describe in Section \ref{subtract_sec} a subtraction technique that we use in our data analysis.  In Section \ref{gain_sims}, we discuss the amplification of short-duration seed pulses in PIC simulations and as calculated by a coupled-mode solver.  Section \ref{late_time} covers our observation of kinetic inflation that occurs after the seed pulse passes, and Section \ref{measure_inflation} covers our measurements of kinetic inflation using longer duration seeds.  Finally, we discuss the onset of inflation with continuous seeds in Section \ref{continuous} and conclude in Section \ref{conclusion}.

\section{Geometry and Plasma Conditions}\label{geom}

\begin{figure}[tbp]
\includegraphics[width=\linewidth]{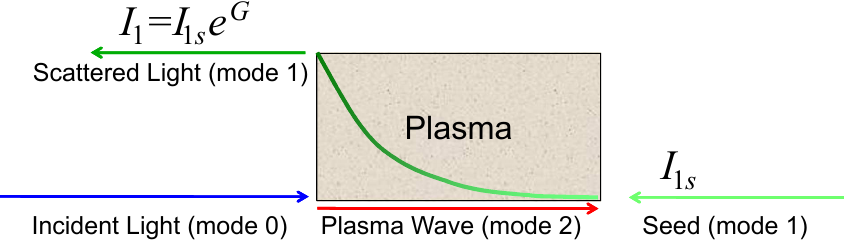}
\caption{The geometry of the OSIRIS simulations.\label{geometry}}
\end{figure}
Throughout this paper, we use normalized units to describe the OSIRIS simulations. To make connection to parameters of interest for ICF, we assume that the incident pump has a  wavelength of $\lambda_0=$351nm.  When we note quantities in physical units, they correspond to this pump wavelength. Our formulas and other quantities are given in CGS units while the temperature is often given in units of eV.

Figure \ref{geometry} depicts the usual simulation geometry.  The pump laser ($\omega_0, \vec{k}_0$) is incident from the left with a normalized electric field amplitude $E_0e/m_ec\omega_0=3.68\times10^{-3}$ (in physical units, $I_0=1.5\times10^{14}$ W/cm${}^2$).  The seed ($\omega_1, \vec{k}_1$) enters the simulation box from the right and beats with the pump, inducing an electrostatic wave ($\omega_2, \vec{k}_2$) in the plasma, which travels to the right.  Light from the pump scatters off the plasma wave and amplifies the seed.  For the simulation parameters, this process is a convective instability.

Our simulation plasma has fixed ions and we do not add collisions.  We simulate densities and temperatures for which SRS is expected to occur for ignition experiments at the National Ignition Facility (NIF).\cite{moses:nif}  The plasma has a uniform density $n=0.12n_c$ ($\approx1.1\times10^{21}$ cm${}^{-3}$), where $n_c$ is the critical density for the pump laser.  The electron thermal speed is $v_{th}=0.0699c$ ($T_e=2.5$ keV).  The box has length $L=1790c/\omega_0$ (100 $\mathrm{\mu m}$) and 8192 cells, giving a cell width of 1.1$\lambda_D$, where $\lambda_D$ is the Debye length.  We use 16,384 particles per cell and quadratic splines for the  particle shape to reduce spurious noise due to aliasing.  The particle boundary conditions are thermalizing and we use perfectly matched layers\cite{berenger:perfectly} for the field boundary conditions.  The particle pusher is relativistically correct in all the simulations except where we state otherwise.

The seed intensities we use in our simulations are far above the
background electromagnetic noise in ICF, which for BSRS is primarily
Thomson scattering.\cite{Strozzi:Ray}  For the plasma conditions used in
our simulations and a typical NIF quad (four laser beams arranged in a
square) with effective F-number\cite{moses:nif} of 8 as the pump, the
effective Thomson scattering seed within the FWHM of the peak relativistic
gain (discussed below) is $1.6\times10^4$ W/cm${}^2=1.1\times10^{-10}I_0$.
 SRS growth from such noise is initially linear, and enhanced over
plane-wave growth by intense speckles in a phase-plate-smoothed beam. In
down-stream regions, this light acts as an SRS seed far above thermal
noise, and may reach amplitudes where kinetic effects are significant.  We
choose seed intensities to induce such effects.

\section{Linear Theory of Convective BSRS Gain\label{gain}}
\subsection{Non-Relativistic Theory}\label{nonrel_gain}
Our PIC simulations never reach a steady state.  However, linear theory states that the reflectivity will quickly reach a steady state when we use a continuous scattered light wave seed.  In the convective steady state, the seed intensity is amplified by a factor of $e^G$ by the time it exits the box.  $G$ is the linear intensity gain exponent, commonly called ``the gain," and we present an equation for it in the strong damping limit.  Here, we summarize the results from a detailed derivation for the steady-state gain from Ref. \onlinecite{Strozzi:Ray}.

Given the pump ($\omega_0$, $\vec{k}_0$) and the seed ($\omega_1$, $\vec{k}_1$), we calculate the plasma wave ($\omega_2$, $\vec{k}_2$) using the matching conditions,
\begin{subequations}
\begin{equation}
\omega_0 = \omega_1 + \omega_2 \label{w_cond}
\end{equation}
and
\begin{equation}
\vec{k}_0 = \vec{k}_1 + \vec{k}_2. \label{k_cond}
\end{equation}
\end{subequations}
We need $\omega_2$ and $\vec{k}_2$ for the plasma susceptibility when we calculate the gain.

Let the seed intensity be denoted by $I_1(z)$, where $z=(0, L)$ is the (left, right) edge of the box.  Then,
\begin{equation}
I_1(z)=e^{G_l(z)}I_1(L),\label{convective_gain}
\end{equation}
where $G_l(z)$ is the linear intensity gain exponent,
\begin{equation}
G_l(z)\equiv \int_z^L \Gamma_1(z')I_0(z')dz'.\label{gain_eq}
\end{equation}
We neglect pump depletion and light wave damping, so $I_0$ is constant, which leads to
\begin{equation}
\Gamma_1 \equiv \Gamma_s \mathrm{Im}\left[\frac{\chi_e}{\varepsilon}(1+\chi_I)\right], \label{gamma1}
\end{equation}
where the subscripts $e$ and $I$ denote the electron and ion species, respectively, $\chi_j$ is the collisionless susceptibility for species $j$, $\varepsilon(k_2,\omega_2) = 1+\sum_j \chi_j(k_2,\omega_2)$ is the plasma dielectric function, and

\begin{equation}
\Gamma_s \equiv \frac{2\pi r_e}{m_ec^2}\frac{1}{\omega_0}\frac{k_2^2}{k_0|k_1|},\label{gamma_s}
\end{equation}
where $r_e\equiv e^2/m_ec^2$ is the classical electron radius.  For a Maxwellian velocity distribution, $\chi_j$ is given by
\begin{equation}
\chi_j(k_2,\omega_2) = - \frac{\omega_{pj}^2}{2k_2^2v_{Tj}^2}Z'\left(\frac{\omega_2}{\sqrt{2}k_2v_{Tj}}\right),
\end{equation}
where $\omega_{pj}$ is the plasma frequency of species $j$, $v_{Tj} = \sqrt{T_j/m_j}$ is the thermal speed of species $j$, and $Z'(s) = dZ/ds.$  $Z'(s)$ must be calculated numerically, and is typically found by first computing $Z(s),$ the plasma dispersion function.\cite{fried:plasma}  $Z'(s)=-2sZ(s)-2,$ with \begin{equation}
Z(s) = i\sqrt{\pi}e^{-s^2}[1+\mathrm{erf}(is)].
\end{equation}
$\chi_j\to-(\omega_{pj}/\omega_2)^2$ as $m_j\to\infty$, so we can set $1+\chi_I\to1$ everywhere (recall we use fixed ions in the OSIRIS simulations).  In particular, $\varepsilon=1+\chi_e$ and
\begin{equation}
\Gamma_1 = \Gamma_s \mathrm{Im}\left[\frac{\chi_e}{\varepsilon}\right].\label{gamma1_simple}
\end{equation}

We further simplify Eq. \ref{gain_eq} since we are dealing with a uniform plasma.  $\Gamma_1$ is constant, so the gain is given by
\begin{equation}G_l(z) = \Gamma_1I_0(L-z).\end{equation}
We also define an amplitude gain rate, $g_0$, as
\begin{equation}g_0 = \frac{\Gamma_1I_0}{2}\propto \frac{\chi_i}{(1+\chi_r)^2 + \chi_i^2},\label{g0}\end{equation}
where $\chi_r$ and $\chi_i$ are the real and imaginary parts of $\chi$, respectively.  We plot the theoretical gain spectrum for the conditions of interest in Figure \ref{gain_curves} as dash-dotted lines.

Equations \ref{convective_gain}-\ref{gamma_s} are valid in the strong damping limit.  This limit applies when $|v_{g2}\partial a_2/\partial x|\ll|\nu_2a_2|$, where $v_{g2}$ is the plasma wave group velocity, $a_2$ is the plasma wave action amplitude (defined in Section \ref{CMEs}), and $\nu_2$ is the Landau damping\cite{Landau:vibrations} rate.  In a homogeneous plasma in the convective steady-state, which is where our gain calculation applies, this condition is $g_0\ll\nu_2/v_{g2}$.  Working at the peak of the non-relativistic gain curve, we have a spatial gain rate of $g_0^{nr}=3.28\times10^{-4}\omega_0/c$ and $\nu_2/v_{g2}=0.0611\omega_0/c$.  Therefore, we are in the strong damping limit.

\subsection{Relativistic Modification}\label{rel_mod}
We now explore the impact of special relativity on linear gain.  Estabrook and Kruer\cite{Estabrook:Theory} included an analysis of  SRS for temperatures for which relativistic corrections are important, and performed 1.5D PIC simulations of  laser and plasma conditions where  the plasma wave is weakly damped.  They found that non-relativistic linear theory does not adequately describe the wavenumber of the fastest growing mode in high-temperature ($\sim$64 keV) plasmas, but taking into account the effective (reduced) plasma frequency and corresponding density due to special relativity bringsƒ theory and simulation into better agreement.  More recently, Bergman and Eliasson derived a fully relativistic expression for the unmagnetized plasma dielectric function,\cite{bergman:linear} and Bers et.\ al.\ derived approximate expressions relevant to current and near-future deuterium-tritium fusion plasmas.\cite{bers:relativistic}  Palastro et. al. have also derived a fully relativistic description of Thomson scattering.\cite{palastro:fully}

In this subsection, we simply make some heuristic changes to the formulas in the previous subsection to account for special relativity.  We replace the susceptibility in Eq.\ \ref{gamma1_simple} with the relativistic one of Bergman and Eliasson, which is computed using a 3D J\"uttner-Synge distribution, as opposed to a Maxwellian distribution.

Given the 3D J\"uttner-Synge distribution,
\begin{equation}
f(\gamma) = \frac{n_0\mu e^{-\mu\gamma}}{4\pi m_e^3c^3K_2(\mu)},
\end{equation} the electron susceptibility is given by 
\begin{equation}
\chi_e(\kappa_2,\Omega_2)=\frac{\mu}{\kappa_2^2}\left[1-\frac{\mu}{K_2(\mu)}\frac{\partial^2}{\partial\mu^2}\frac{P(\mu,\kappa_2/\Omega_2)}{\mu}\right],
\end{equation}
where $\gamma$ is the relativistic factor, $\mu\equiv m_ec^2/T_e$, $K_2(\mu)$ is the modified Bessel function of the second kind, $\Omega_2 \equiv \omega_2/\omega_{pe}$, and $\kappa_2 \equiv k_2c/\omega_{pe}$.  $P$ is given by
\begin{eqnarray}
P(\mu,\kappa_2/\Omega_2) \equiv && \int_1^\infty \frac{e^{-\mu\gamma}}{\sqrt{\gamma^2-1}}\frac{d\gamma}{\gamma^2(1-\kappa_2^2/\Omega_2^2)+\kappa_2^2/\Omega_2^2}\nonumber\\
&&-\frac{i\pi\sigma}{2\kappa_2/\Omega_2}e^{-\mu (\kappa_2/\Omega_2) / \sqrt{\kappa_2^2/\Omega_2^2-1}}\label{Pfun},
\end{eqnarray}
where $\sigma=0$ for $\Omega_2^2\geq\kappa_2^2$ and $\sigma=1$ for $\Omega_2^2<\kappa_2^2$.

We can also take into account relativistic effects in the electromagnetic dispersion relation, $\omega^2=\omega_{pe}^2+c^2k^2,$ by using a relativistic version of the plasma frequency:
\begin{eqnarray}\omega_{pe}^2 \to && \omega_{pe}^2\frac{\mu^2}{K_2(\mu)}\int_1^\infty \frac{\partial^2}{\partial\mu^2}\left(\frac{e^{-\mu\gamma}}{\mu}\right)\frac{\sqrt{\gamma^2-1}}{\gamma^4} d\gamma\\
&&\approx \omega_{pe}^2\left(1-\frac{5}{2\mu}\right)~~\textrm{for}~\mu \gg 1.\nonumber \end{eqnarray}
We use this relativistic plasma frequency when we calculate $k_0$ and $k_1$.  The decrease in the effective plasma frequency with temperature is due to relativistic corrections to the internal energy of the plasma.\cite{tzeng:suppression}  However, this change only has a relatively small impact on the gain at the temperature we use in our simulations.  It is also possible to calculate $k_0$ and $k_1$ using the fully relativistic transverse dispersion relation from Ref.\ \onlinecite{bergman:linear}, but we did not attempt to do so.

The overall effect of special relativity on the gain curve is shown in Figure \ref{gain_curves}.  Notice that the peak of the relativistic gain curve lies above the peak of the non-relativistic gain curve, and the peak occurs at a shorter wavelength.  This difference occurs because there are less particles and a shallower slope at the plasma wave's phase velocity in the J\"uttner distribution than in the Maxwellian distribution, resulting in weaker Landau damping.  At the peak of the analytic non-relativistic gain curve ($\lambda_1 = 1.659\lambda_0 = 582.37$ nm), for which $k_2\lambda_{De}=0.289$, the non-relativistic damping rate is $3.24\times10^{-3}\omega_0$, while the relativistic rate is $2.44\times10^{-3}\omega_0.$ The strong damping limit still applies at the peak of the relativistic gain curve ($\lambda_1 = 1.655\lambda_0 = 580.88$ nm), for which the spatial gain rate is $g_0^r=4.43\times10^{-4}\omega_0/c$ and $\nu_2/v_{g2}=0.0465\omega_0/c$.

We also performed some gain calculations using the approximate expression of Bers et.\ al.\ for the relativistic longitudinal dispersion relation.  Their expression shifts the gain curve down in wavelength significantly more than numerically integrating the formula of Bergman and Eliasson.

\subsection{PIC Modification}\label{pic_mod}
We can improve the agreement between simulation and theoretical results by taking into account a few known aspects of finite-difference PIC codes: finite-size particles, differencing operators, and field smoothing plus compensation.\cite{birdsall:plasma}   The particles have a finite size because the charge and current are interpolated to a grid via the ``shape factor" $S(\vec{x})$: 
\begin{equation}q\delta(\vec{x}) \to qS(\vec{x}).\end{equation}
To reduce the self-heating and spurious noise from aliasing, we use second-order B-splines. Transforming to Fourier space,
\begin{equation}qS(\vec{x}) \to qS(\vec{k}),\end{equation}
where
\begin{equation}S(k) = \frac{1}{L}\left(\frac{\sin(k\Delta/2)}{k\Delta/2}\right)^3\end{equation}
for 1D simulations, with $\Delta$ being the cell width.

In finite-difference codes, like OSIRIS, differencing operators modify the dispersion relation by changing the relationship between the charge density $\rho$, longitudinal electric field $E_2$, and electrostatic potential $\phi$.  In Fourier space,
\begin{eqnarray}
4\pi\rho(k_2) &&= k_2^2\left(\frac{\sin(k_2\Delta/2)}{k_2\Delta/2}\right)^2\phi(k_2)\nonumber\\
&&= K^2(k_2)\phi(k_2)
\end{eqnarray}
and
\begin{eqnarray}
E_2(k_2) &&= -ik_2\frac{\sin(k_2\Delta)}{k_2\Delta}\phi(k_2)\nonumber\\
&&= -i\kappa(k_2)\phi(k_2).
\end{eqnarray}

We additionally smooth the fields in our simulations to further reduce the effects of aliasing, and we compensate to reduce numerical modifications to the dispersion relation for small $\vec{k}$.  Without the use of splines and smoothing, grid heating instabilities occur for $\Delta \gtrsim 3\lambda_{De}$.  Finite-difference codes, such as OSIRIS, can use a digital filter to compensate for this effect.  We perform the filtering of some quantity $\phi$ on the grid by replacing
\begin{equation}\phi_j \mathrm{~with~} \frac{W\phi_{j-1}+\phi_j+W\phi_{j+1}}{1+2W},\end{equation}
where $j$ is the grid index and $W$ is a weighting factor.  Transforming into Fourier space,
\begin{eqnarray}
\phi_f(k) &&= \frac{1+2W\cos(k\Delta)}{1+2W}\phi_0(k)\nonumber\\
&&= SM_W(k\Delta)\phi_0(k).
\end{eqnarray}
We use two filters in the simulations in this paper. The first is a two-pass filter, which we use unless stated otherwise.  The first pass has a stencil of $\frac{1}{4}$(1,2,1) (W=1/2) and the second has a stencil of $\frac{1}{4}$(-1,6,-1) (W=-1/6).  Therefore,
\begin{eqnarray}
SM(k)&&=SM_{1/2}(k)SM_{-1/6}(k)\nonumber\\
&&=\frac{1+\cos(k\Delta)}{2}\frac{3-\cos(k\Delta)}{2}.
\end{eqnarray}
The second filter has five passes, and we choose it because it causes less deviation from the longitudinal dispersion relation without PIC effects than the 2-pass filter.  The first four passes use a stencil of $\frac{1}{4}$(1,2,1) (W=1/2) and the last pass uses a stencil of $\frac{1}{4}$(-5,14,-5) (W=-5/14).  We perform additional simulations with this filter to observe the effect on the gain curve.  
\begin{eqnarray}
SM(k)&&=SM_{1/2}^4(k)SM_{-5/14}(k)\nonumber\\
&&=\left(\frac{1+\cos(k\Delta)}{2}\right)^4\frac{14-10\cos(k\Delta)}{4}.
\end{eqnarray}

The particle shape factor, differencing operators, and field smoothing only affect the plasma frequency.  We simply make the change
\begin{equation}\omega_{pe}^2 \to \omega_{pe}^2\frac{k_2\kappa(k_2)}{K^2(k_2)}(L\cdot S(k_2))^2SM(k_2)\end{equation}
everywhere $\omega_{pe}$ appears in our formulas to account for their effects.

\begin{figure}
\includegraphics[width=\linewidth]{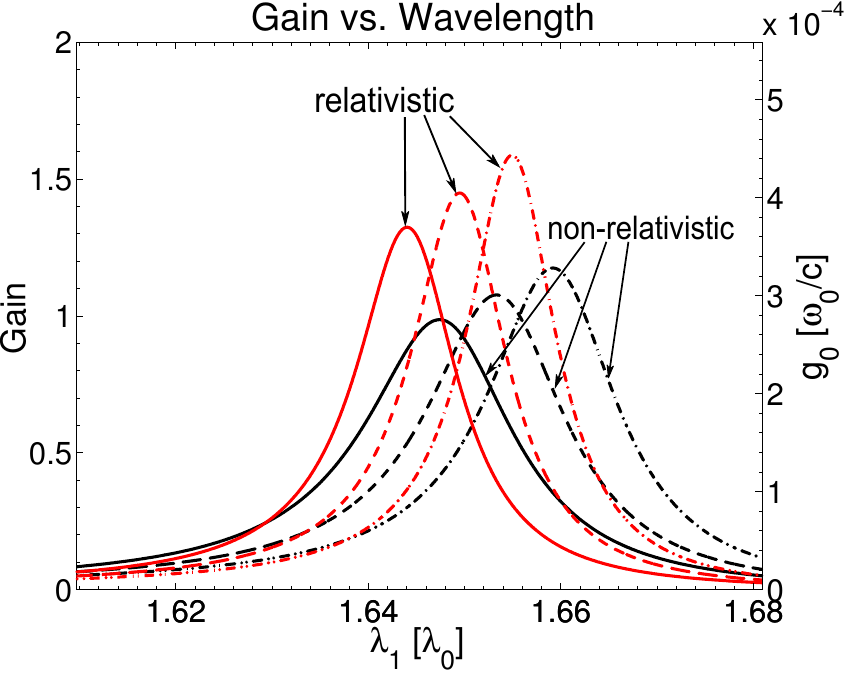}
\caption{Several calculations of the linear convective BSRS gain spectra for a box $1790c/\omega_0$ (100 $\mathrm{\mu m}$) long.  The dash-dotted curves are analytic results, and the (dashed, solid) ones take into account PIC effects with the (5,2)-pass filter.\label{gain_curves}}
\end{figure}

The effect of the shape factor, differencing operators, and both the 2-pass and 5-pass filters on the relativistic and non-relativistic gain curves is shown in Figure \ref{gain_curves}.  Checking the strong damping limit condition for the 2-pass filter, in the non-relativistic case, the peak gain rate drops to $g_0^{nr}=2.75\times10^{-4}\omega_0/c$ with $\nu_2/v_{g2}=0.0636\omega_0/c$, while for the relativistic case, it drops to $g_0^r=3.70\times10^{-4}\omega_0/c$ with $\nu_2/v_{g2}=0.0531\omega_0/c$.

\section{Subtraction Technique}\label{subtract_sec}
\begin{figure}[tbp]
\includegraphics[width=\linewidth]{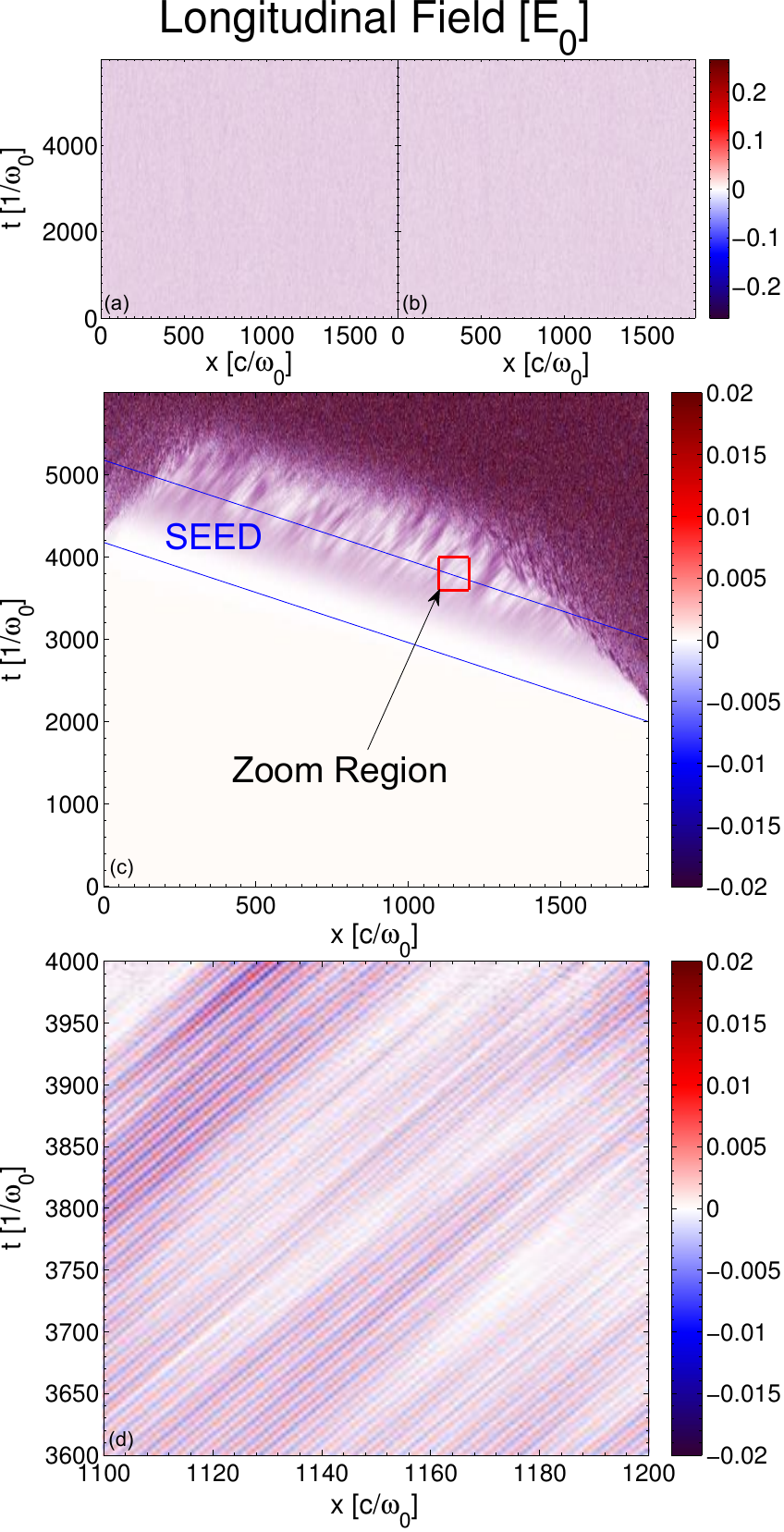}
\caption{The longitudinal electric field in a simulation with (a) and without (b) a seed pulse.  Subtracting the latter from the former reveals the plasma wave (c,d).\label{subtract}}
\end{figure}

We use a subtraction technique\cite{decyk:simulation} in our data analysis to clearly see waves with amplitudes below the background plasma fluctuation level.  The technique requires running two simulations, the first with a perturbation whose effects we wish to examine, and the second without the perturbation, but with the same random number generator seed.  We then subtract the results of the second simulation from the results of the first.  In our case, the first simulation has both a backward propagating light seed pulse and a forward propagating pump, while the second simulation has just the pump.

Figure \ref{subtract} shows the electrostatic field induced by the beating of the pump and the seed pulse.  The seed pulse in this simulation has a Gaussian-like profile, as described in section \ref{gain_sims}, with $\lambda_1= 1.644\lambda_0$ ($577$nm) and $I_{1s}=5\times10^{-4}I_0$.  The amplitude of the plasma wave is so small that we cannot distinguish it from the background fluctuations without using the subtraction technique.  In the subtracted result, background fluctuations enter the simulation starting at the sides of the box due to the thermalizing boundary conditions.  Fluctuations will always re-enter the subtracted data as the two simulations become uncorrelated, but the re-emission of particles with random speeds at the boundaries exacerbates this situation.

Because we do not observe meaningful SRS without a seed, we use the subtraction technique as a means of separating the scattered light from the pump light when pump depletion is not significant.  When pump depletion is small, the subtraction technique for the transverse electric field works well for finding the scattered light at all positions in the box.  However, when pump depletion becomes significant, the subtraction technique does not produce good results by itself anywhere except at the far left side of the box, where pump depletion does not occur.  Therefore, we can still use the subtraction technique to observe the reflectivity at the left side of the box, but we need to filter out the pump in Fourier space to observe the scattered light anywhere else in the box.

The number of particles per cell in the simulation affects the usefulness of the subtraction technique.  Fluctuations take longer to enter the simulation as we increase the number of particles per cell.  When we decrease the number of particles per cell, low-intensity seed pulses become more difficult to distinguish from the background fluctuations when we use the subtraction technique, until we cannot distinguish the peak of a pulse with $I_{1s}=5\times10^{-4}I_0$ from the fluctuations in simulations with 512 particles per cell.  However, changing the number of particles per cell has no significant effect on the convective amplification of the seed pulse.

\section{Convective Gain Simulations \& Coupled-Mode Results\label{gain_sims}}
\subsection{OSIRIS Convective Gain Simulations}
\begin{figure}[tbp]
	\includegraphics[width=\linewidth]{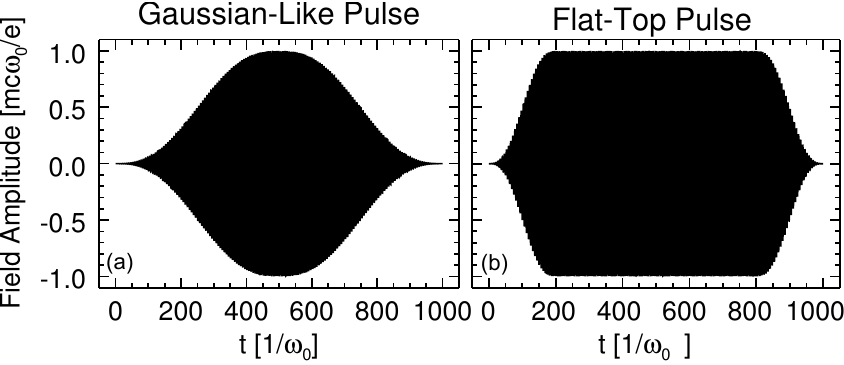}
	\caption{The shapes of the seed pulses in the simulations.  A Gaussian-like pulse (a) with a 500$\omega_0^{-1}$ rise and fall time ($\sim500\omega_0^{-1}$ FWHM), and a flat-top pulse (b) with a 200$\omega_0^{-1}$ Gaussian-like rise and fall, and a steady amplitude for 600$\omega_0^{-1}$ in between.\label{pulse_shapes}}
\end{figure}

We perform OSIRIS PIC simulations to observe seed amplification in the linear regime and determine under what conditions SRS enters the nonlinear regime.  The simulations begin at $t=0$ with the pump incident from the left.  After $2,000\omega_0^{-1}$, the seed enters from the right, by which time the pump has crossed the box.  We use two different temporal profiles, or ``shapes,'' for the seed and vary its wavelength and intensity.  The first shape is approximately Gaussian, and rises from zero to its peak amplitude in $\tau=500\omega_0^{-1}$, then falls back to zero over another $500\omega_0^{-1}$, for $\sim500\omega_0^{-1}$ FWHM.  The second shape is a flat-top pulse with a Gaussian-like rise and fall time of $\tau=200\omega_0^{-1}$, and a constant peak amplitude for 600$\omega_0^{-1}$ in between.  The two pulse shapes are plotted in Figure \ref{pulse_shapes}.  We describe the seed pulse using the notation $I_1(z=L,t)=I_{1s}s(t)$, so that $s(t)$ describes the pulse shape and $I_{1s}$ is the maximum incident intensity.  In our simulations using seed pulses, $I_{1s}\geq5\times10^{-4}I_0$.

\begin{figure}[tbp]
   \includegraphics[width=0.96\linewidth]{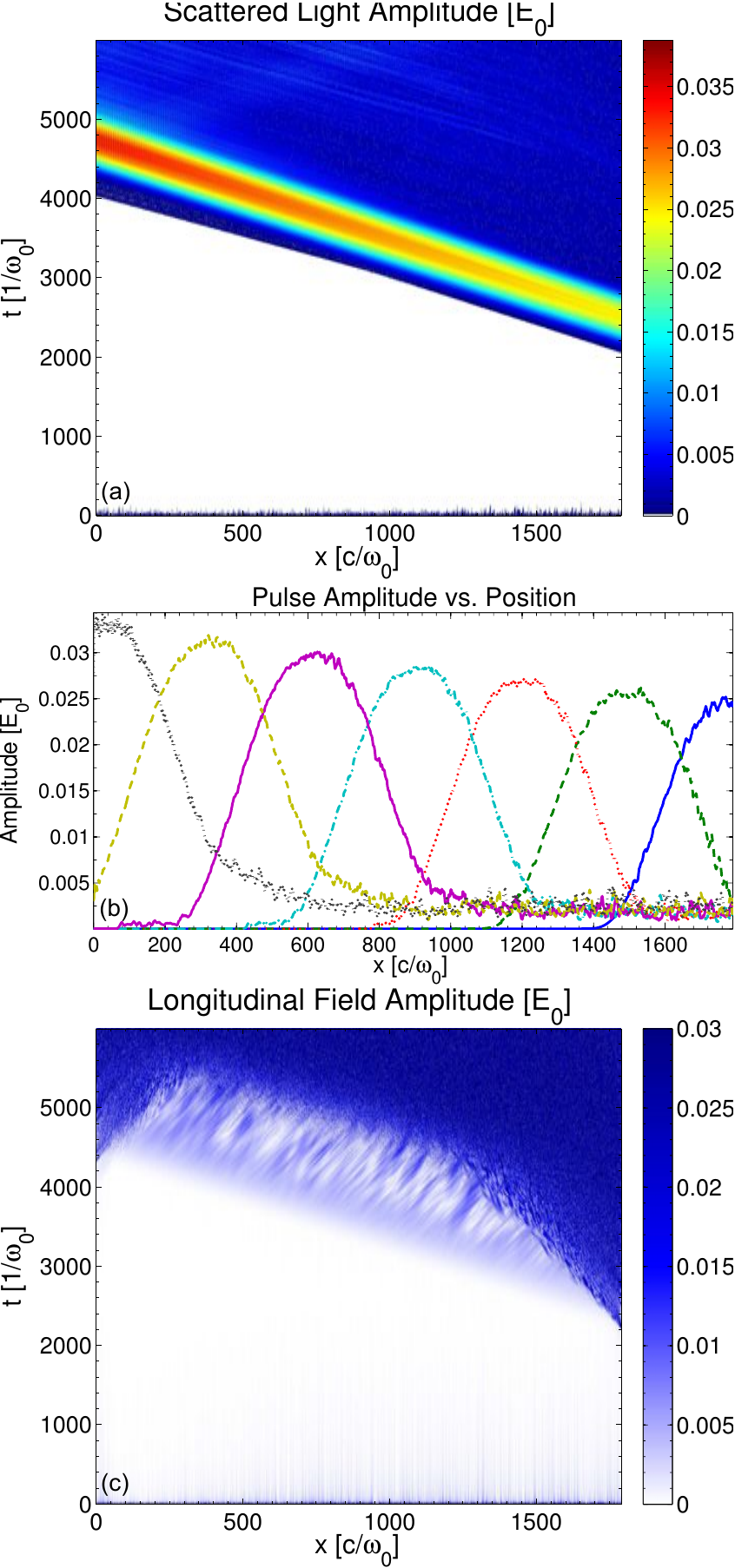}
	\caption{Scattered light vs.\ position and time (a), lineouts of the scattered light vs.\ position at various times (b), and plasma wave amplitude (c) from a simulation using a Gaussian-like seed pulse with $I_{1s} = 5\times10^{-4}I_0$ and $\lambda_1=1.644\lambda_0$.\label{EM_EPW}}
\end{figure}

Figure \ref{EM_EPW} shows the scattered light and the plasma wave as a function of position and time in an OSIRIS PIC simulation when we use a Gaussian-like seed pulse with $I_{1s}=5\times10^{-4}I_0$ and $\lambda_1 = 1.644\lambda_0$.  We also include line-outs of the scattered light amplitude vs. position at various times to show the evolution of the seed pulse more clearly as it crosses  the box from right to left.  We use a Hilbert transform to envelope the results, producing a smooth appearance.

\begin{figure}[tbp]
	\includegraphics[width=\linewidth]{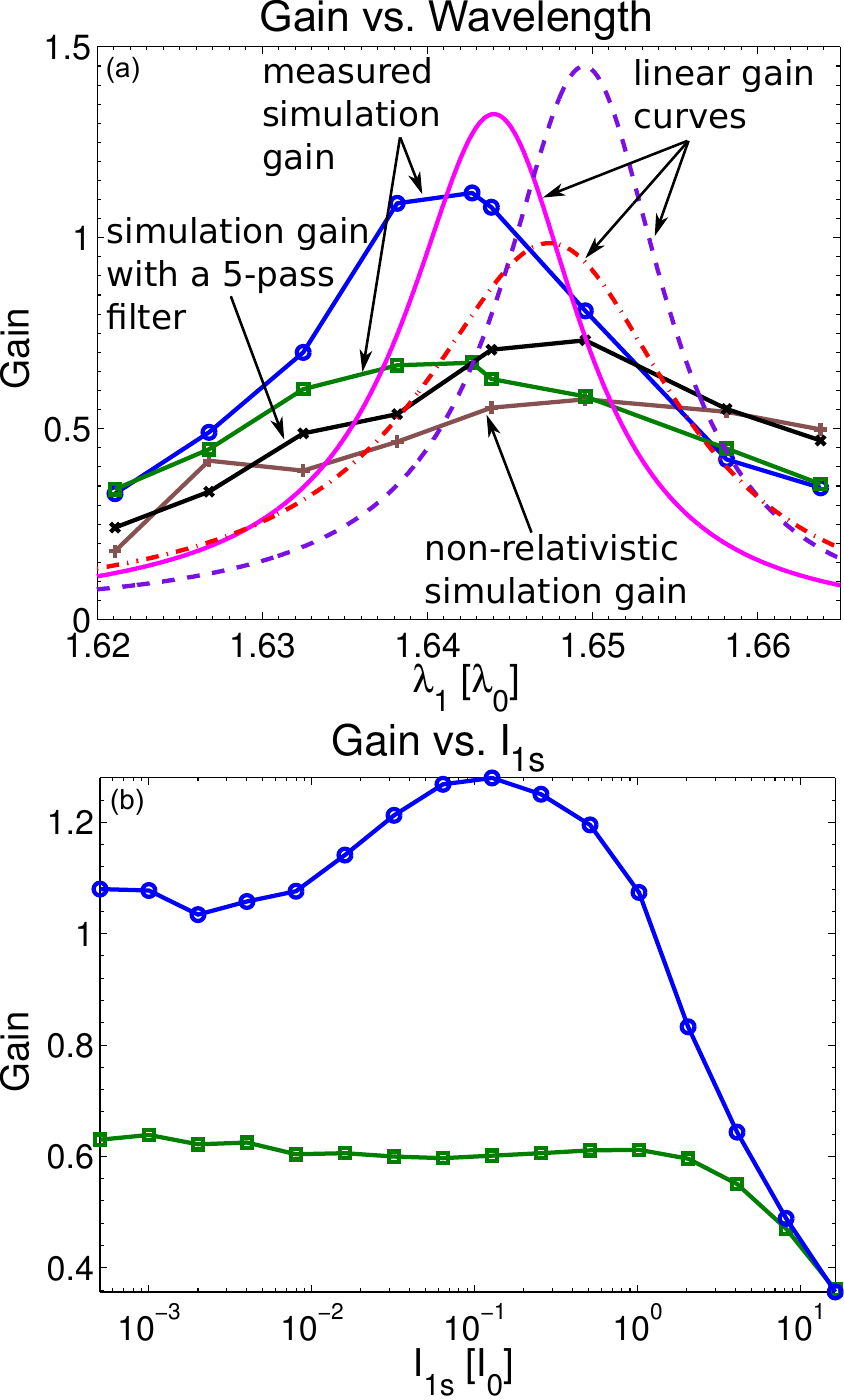}
	\caption{Measured gain in simulations with box length $1790c/\omega_0$ (100 $\mathrm{\mu m}$) vs. seed wavelength (a) and seed intensity $I_{1s}$ (b).  Plotted are the gains we measure using a Gaussian-like pulse (green with square markers) and a flat-top pulse (blue with circle markers), both with $\lambda_1=1.644\lambda_0$.  (a) includes the gain from simulations with a Gaussian-like pulse and the 5-pass filter (black with `x' markers), plus those without relativistic effects (brown with `+' markers).  Several theoretical gain curves taking into account PIC effects are included: 2-pass non-relativistic is dash-dotted red, 2-pass relativistic is solid magenta, and 5-pass relativistic is dashed purple.\label{sim_gain}}
\end{figure}

We define the measured gain, $g_{meas}$, of a pulse as
\begin{equation}
	g_{meas} \equiv \ln\left(\frac{\textrm{max}(I_1(z=0,t))}{I_{1s}}\right).
\end{equation}
The plot at the top of Figure \ref{sim_gain} shows the gain we measure in simulations when we vary the seed wavelength while keeping $I_{1s}$ fixed at $5\times10^{-4}I_0$.  The linear relativistic gain peaks near $\lambda_1=1.644\lambda_0$ ($577$nm) in simulations with a 2-pass filter and near $\lambda_1= 1.650\lambda_0$ ($579$nm) in simulations with a 5-pass filter.  Our simulation results agree with these predictions.  We also plot a gain curve from simulations using a non-relativistic particle pusher and a Maxwellian velocity distribution.  This non-relativistic curve lies below the relativistic one at most points, as we expect, and the location of its peak agrees with the gain curve from non-relativistic theory with a 2-pass filter.

In the plot on the bottom of Figure \ref{sim_gain}, we see how the measured gain changes as we vary $I_{1s}$ while keeping the seed wavelength fixed at $\lambda_1=1.644\lambda_0$.  The measured gain of the Gaussian-like pulse remains relatively constant as we increase the initial seed intensity, until the seed intensity reaches several times the pump intensity.  This behavior indicates that we are in the linear regime for even large seed amplitudes.  The deviation from linear theory at the highest amplitudes is due to pump depletion.  However, the measured gain of the flat-top pulse increases with seed intensity before falling off.  This inflationary gain is caused by the higher amplitude seed pulse generating a larger amplitude plasma wave and, in the presence of the seed pulse, particles executing several bounces, thus decreasing the Landau damping rate.  We explain this effect in more detail in the following sections.  We note that if we had decreased (increased) the seed pulse length, the deviation from linear behavior would occur at higher (lower) seed intensity.

\subsection{The Coupled-Mode Equations}\label{CMEs}
The measured gain of a seed pulse can differ from the steady-state linear result due to several linear and nonlinear effects. Linear effects include pulse shape, with each frequency in a spectrum of incident frequencies being amplified at a different rate.  Nonlinear effects include pump depletion and nonlinear (kinetic) and nonlocal reductions to the real part of the frequency and the damping rate of the plasma wave.   In this section, we investigate the effect of the pulse shape on the measured gain using the coupled-mode equations.\cite{berger:dominant,strozzi:dissertation}  A comparison of the coupled-mode and OSIRIS results, similar to that performed by Wang et.\ al.\ in Ref.\ \onlinecite{wang:feasibility}, isolates linear from truly nonlinear, kinetic physics, and provides confidence in the PIC method.

In the coupled-mode equations, we let ($\omega_i$, $\vec{k}_i$) of the carrier waves be real and work with complex envelopes $a_i(\vec{x},t)$.  We assume that the envelopes vary slowly with respect to the carriers, such that $|\nabla a_i|\ll|\vec{k}_ia_i|$ and $|\partial a_i/\partial t|\ll|\omega_ia_i|$.  The complex envelopes for the action amplitudes $a_j$ relate to the physical quantities by
\begin{equation}
\vec{A}_j = -i\left(\frac{2\pi}{\omega_j}\right)^{1/2}a_j \exp[i(\vec{k}_j\cdot\vec{x}-\omega_j t)] \hat{k} + cc,~~~j=0,1
\end{equation}
for light waves, with $A_j$ the vector potential, and by
\begin{equation}
n_1 = \frac{ik_2}{2}\left(\frac{2n_B}{m_e\omega_2}\right)^{1/2}a_2\exp[i(\vec{k}_2\cdot\vec{x}-\omega_2 t)] + cc,
\end{equation}
for the plasma wave.  $n_B$ is the spatially varying background electron density, $n_1$ is the perturbation on top of it, and $cc$ denotes complex conjugate.

The couple mode equations are
\begin{eqnarray}
\left(\frac{\partial}{\partial t} + \vec{v}_{g0}\cdot\nabla + \nu_0 + i\delta_0\right)a_0 &&=Ka_1a_2,\\
\left(\frac{\partial}{\partial t} + \vec{v}_{g1}\cdot\nabla + \nu_1 + i\delta_1\right)a_1 &&=-Ka_0a_2^*,\\
\left(\frac{\partial}{\partial t} + \vec{v}_{g2}\cdot\nabla + \nu_2 + i\delta_2\right)a_2 &&=-Ka_0a_1^*,
\end{eqnarray}
where the coupling constant is
\begin{equation}
K \equiv \frac{k_2}{\sqrt{\omega_0\omega_1\omega_2}}\frac{\omega_{pe}^2}{\sqrt{8n_Bm_e}},
\end{equation}
$\nu_i$ is the damping rate of mode $i$, and $\delta_i \equiv (\omega_{pe}^2-\omega_i^2+c^2k_i^2)/(2\omega_i)$ is the detuning frequency for the light waves, which reflects the departure of mode $i$ from being a natural mode of the plasma. We assume no inverse bremsstrahlung, which is the case in the PIC simulations, so $\nu_0=\nu_1=0$.  The Landau damping rate, $\nu_2=\varepsilon_i/(\partial\varepsilon_r/\partial\omega_2)$, where $\varepsilon_r = \mathrm{Re}[\varepsilon]$, $\varepsilon_i = \mathrm{Im}[\varepsilon]$, and $\varepsilon$ is the kinetic dielectric function.  We set the light-wave detuning, $\delta_0=\delta_1=0$.  For the electrostatic mode, we find $\delta_2$ using the kinetic equation, $\delta_2=-\varepsilon_r/(\partial\varepsilon_r/\partial\omega_2)$.  The equations in Sections \ref{rel_mod} and \ref{pic_mod} allow us to take into account special relativity and PIC effects when we calculate the coefficients.

\begin{figure}[tbp]
	\includegraphics[width=\linewidth]{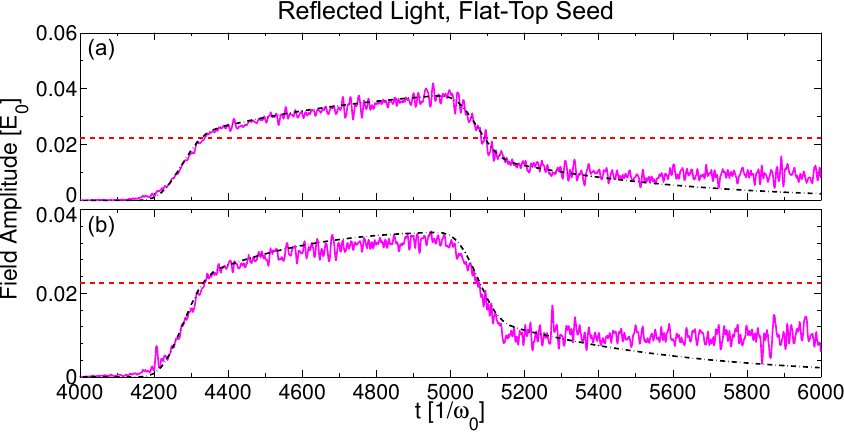}
	\caption{The BSRS reflected light using a flat-top seed with $\lambda_1=1.644\lambda_0$ (a) and $1.650\lambda_0$ (b).  The simulation results are plotted with solid magenta lines while the coupled-mode results are plotted with dash-dotted black lines.  The horizontal dashed red line indicates the maximum seed amplitude with no gain.}\label{CME_reflect}
\end{figure}

\begin{figure}[tbp]
	\includegraphics[width=\linewidth]{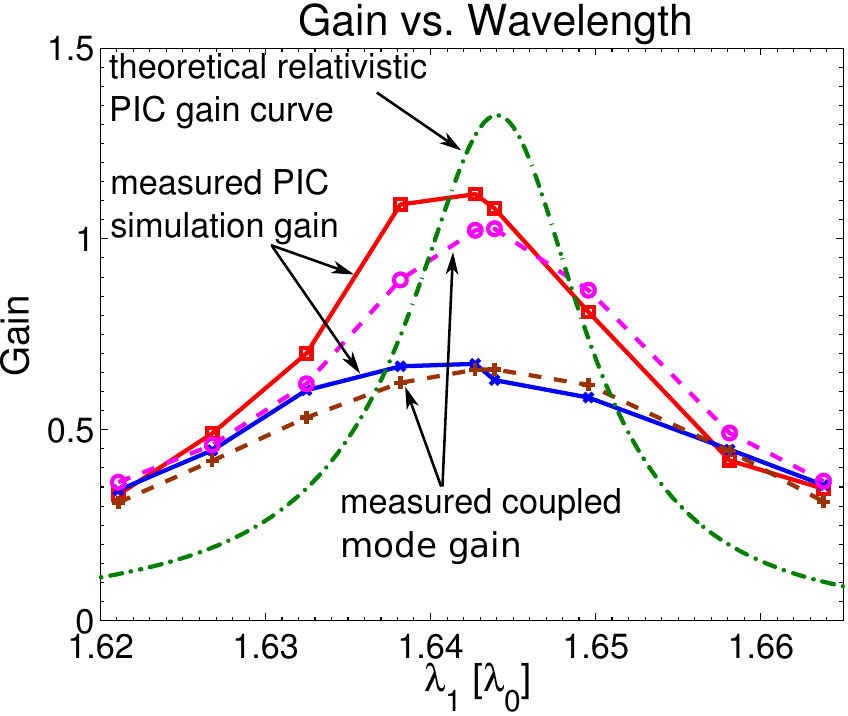}
	\caption{The measured gain in a box of length $1790c/\omega_0$ (100 $\mathrm{\mu m}$) predicted by the coupled-mode solver as we vary the wavelength using a Gaussian-like pulse (dashed brown with `+' markers) and a flat-top pulse (dashed magenta with circle markers).  For comparison, we include the PIC simulation result for the Gaussian-like pulse (solid blue with `x' markers) and the flat-top pulse (solid red with square markers) along with the linear relativistic gain curve taking into account PIC effects (dash-dotted green).}\label{gain_cme}
\end{figure}

We can directly compare the results from OSIRIS and the coupled-mode solver by examining the reflected light.  Figure \ref{CME_reflect} shows the reflected light from runs with $I_{1s}=5\times10^{-4}I_0$ using flat-top pulses with $\lambda_1=1.644\lambda_0$ and $1.650\lambda_0$.  In Figure \ref{gain_cme}, we compare the measured gain from PIC simulations and the coupled-mode solver for various wavelengths.  The simulation and coupled-mode results are in excellent agreement for both the Gaussian-like and flat-top pulse runs.

\begin{figure}[tbp]
	\includegraphics[width=\linewidth]{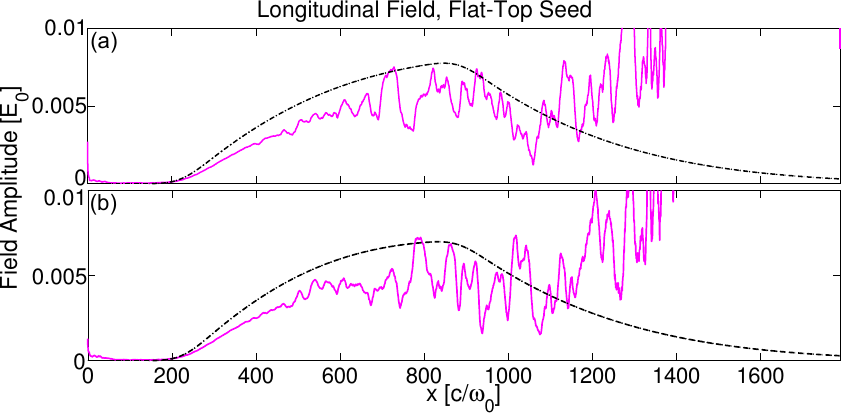}
	\caption{The longitudinal field vs.\ space at $t=4,000\omega_0^{-1}$ when using a flat-top seed with $\lambda_1=1.644\lambda_0$ (a) and $1.650\lambda_0$ (b); the same runs as used in Fig.\ \ref{CME_reflect}.  The simulation results are plotted with solid magenta lines while the coupled-mode results are plotted with dash-dotted black lines.}\label{CME_EPW}
\end{figure}

The agreement between simulation and coupled-mode results is not as good when we examine the longitudinal field.  Figure \ref{CME_EPW} shows the amplitude of the longitudinal field at $t=4,000\omega_0^{-1}$ for the same runs as shown in Figure \ref{CME_reflect}.  The disagreement occurs soon after the wave begins growing and is visible at about $x=300c/\omega_0$.  We are not yet sure of the reason for this disagreement.  The disagreement becomes worse as background fluctuations begin to enter the PIC results after $x=400c/\omega_0$.

\section{Inflation After Seed Passage}\label{late_time}

For short seed pulses, we find significant reflectivity well after the seed propagates out of the simulation box. The pump now directly interacts with the Langmuir wave that is still present in the plasma after the seed leaves the box.  Without trapped particles, the Langmuir wave is described by its linear dispersion relation, BSRS remains in the strongly-damped limit, and no observable growth of BSRS occurs for the pump intensity and the plasma length of interest.  However, a small amplitude wave can evolve into a nonlinear weakly damped wave after the trapped particles execute a few bounces.\cite{oneil:collisionless, morales:nonlinear} The period for a bounce, or bounce time, is $\tau_B\equiv 2\pi/\omega_B$, where $\omega_B=\sqrt{eE_2k/m_e}$ is the bounce frequency for deeply trapped electrons, $E_2$ is the electric field amplitude, and $k$ is the wavenumber of the wave.  As particles are trapped, the damping rate decreases below its linear value to a residual level which depends on details of the problem.\cite{benisti:various}  The ponderomotive beating of the pump and the scattered light will drive the wave to increasing amplitudes.  Such a situation will lead to noticeable reflectivity later in the simulation.  The seed pulse must be the cause of any such reflectivity because, with the pump amplitude we use in our simulations, BSRS is negligible without a seed. BSRS that occurs after the seed has left the box is both useful for isolating the process of kinetic inflation\cite{vu:transient, *vu:kinetic} and is potentially relevant to situations where BSRS in one region of space or time seeds BSRS in another one, creating a plasma wave that triggers an inflationary burst of BSRS.

\begin{figure}[tbp]
\includegraphics[width=\linewidth]{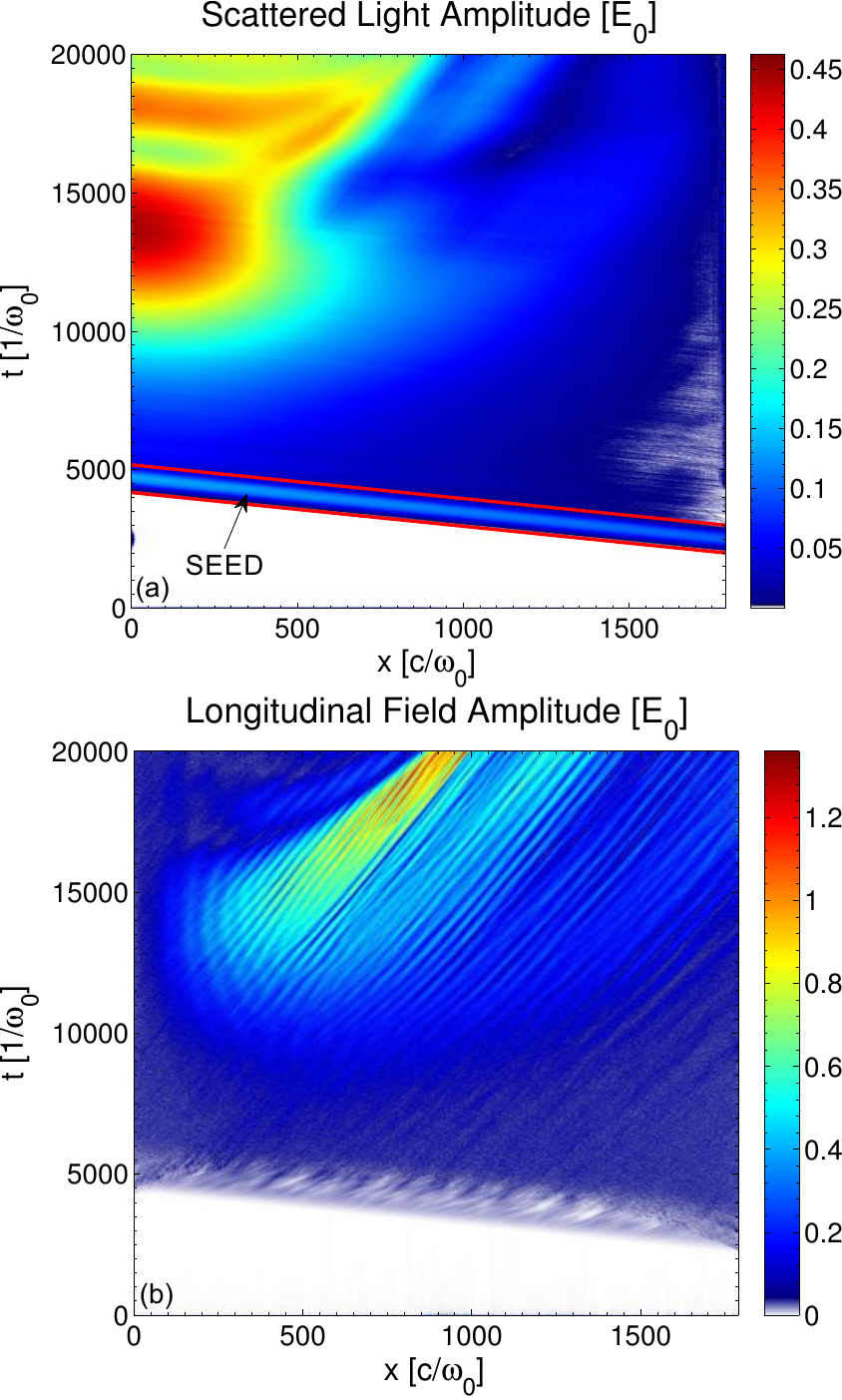}
\caption{The scattered light (a) and the plasma wave (b) seen when we extend the duration of the simulation using a Gaussian-like seed pulse with $I_{1s}=8\times10^{-3}I_0$ and $\lambda_1=1.644\lambda_0$.  The seed exits the box around $t=5,000\omega_0^{-1}$.\label{EM_EPW_lt}}
\end{figure}

\begin{figure}[tbp]
\includegraphics[width=\linewidth]{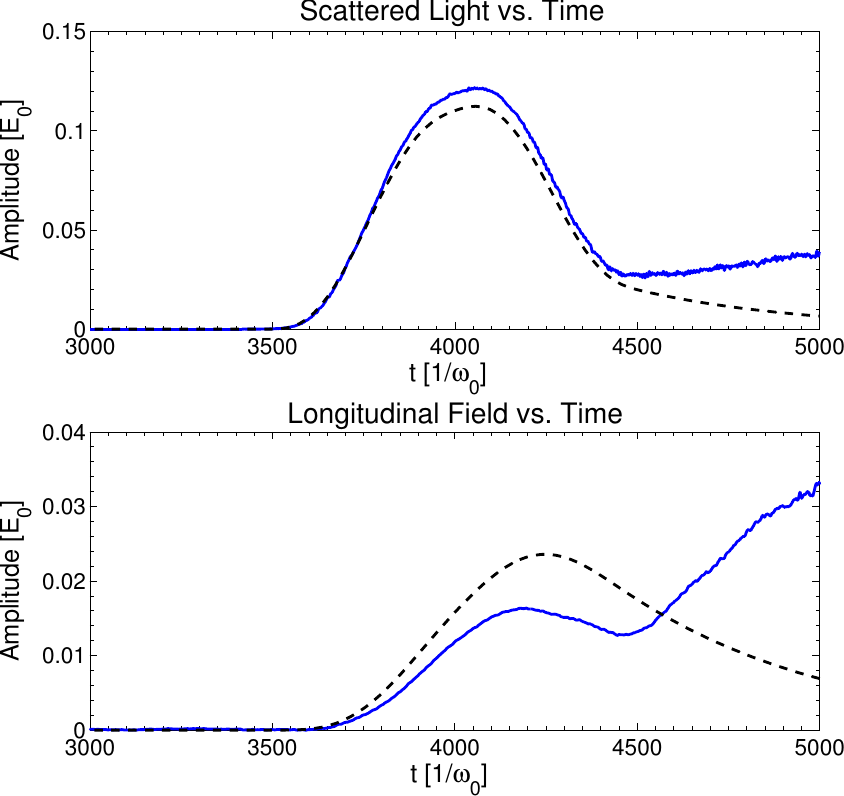}
\caption{Evolution of the scattered light (a) and the longitudinal field (the PIC simulation field is filtered) (b) with time at $x=550c/\omega_0$ for the simulation of Figure \ref{EM_EPW_lt}.  The simulation results are plotted with solid blue lines and the coupled-mode results are plotted with dashed black lines.\label{1.2e12_t_lineouts}}
\end{figure}

Figure \ref{EM_EPW_lt} shows the scattered light and plasma wave for times after the seed leaves the box in a simulation using a Gaussian-like seed pulse with $I_{1s}=8\times10^{-3}I_0$ and $\lambda_1=1.644\lambda_0$.  We observe high reflectivity after $t=10,000\omega_0^{-1}$, along with a corresponding growth in the plasma wave.  Figure \ref{1.2e12_t_lineouts} shows lineouts of the scattered light and longitudinal field at $x=550c/\omega_0$ along with a comparison with the coupled-mode result.  For the longitudinal field from the PIC simulation, we filter out all modes except $1.4\omega_0/c \leq k \leq 1.5\omega_0/c$.  Notice the dip in the plasma wave amplitude around $t=4,500\omega_0^{-1}$, corresponding to the drop in the seed's amplitude, before the plasma wave begins to grow again.

\begin{figure}[tbp]
	\includegraphics[width=\linewidth]{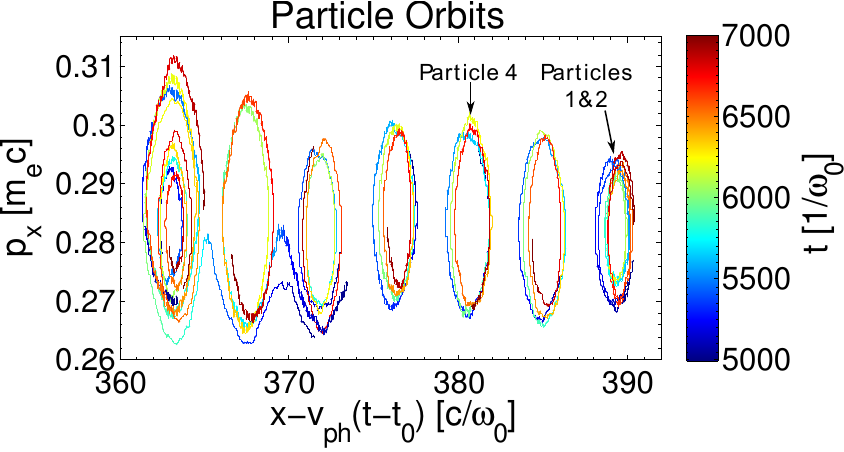}
\caption{Orbits of particles trapped in the plasma wave in the simulation of Figure \ref{EM_EPW_lt}.  The orbits are plotted for $2,000\omega_0^{-1}$ starting at $t_0=5,000\omega_0^{-1}$ in the wave frame, where $v_{ph}=0.271c$ is the plasma wave phase speed and the orbits are centered around the relativistic $p_{ph}=0.282m_ec$.\label{orbits}}
\end{figure}

\begin{figure}[tbp]
	\includegraphics[width=\linewidth]{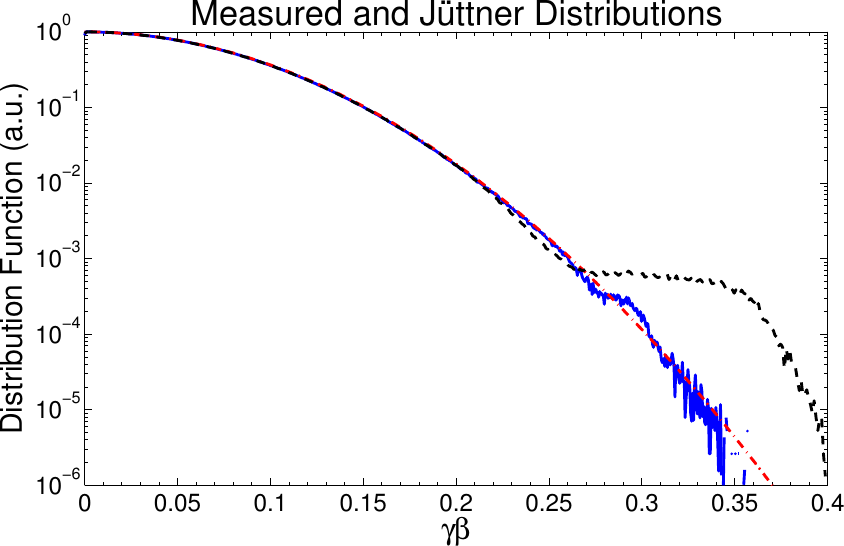}
\caption{The distribution functions from the simulation of Figure \ref{EM_EPW_lt} at $t=6,000\omega_0^{-1}$ (solid blue) and $t=17,000\omega_0^{-1}$ (dashed black) along with the J\"uttner distribution for a 2.5keV electron plasma (dash-dotted red).  The measured distributions have a plateaus beginning around $\gamma\beta=0.26$, indicating particle trapping.\label{distfun}}
\end{figure}

We verify that kinetic inflation is occurring by tracking particles traveling near the plasma wave phase velocity and plotting their orbits in the frame of the wave, as done in Figure \ref{orbits} from $t=5,000\omega_0^{-1}$ to $t=7,000\omega_0^{-1}$.  The elliptic trajectories are clear indicators of particle trapping.  During this time, the distribution function begins to flatten around the Langmuir wave phase velocity, as seen in Figure \ref{distfun}, which is another indication of particle trapping and a clear indication of the reduction of Landau damping.  The tail is flattened to much higher velocities during the larger burst of SRS that grows after the seed leaves; for example, as shown at $t=17,000\omega_0^{-1}$.  This larger tail in our simulations potentially contributes to the energetic electrons (commonly referred to as ``hot electrons") seen in recent experiments.\cite{dewald:hot} The production of hot electron tails by SRS is an active area of research,\cite{winjum:anomalously, yin:trapping} but the small flattening early in time is sufficient to affect the growth and further onset of SRS we are studying here.

Since particle trapping causes the inflationary scattering at late times, the inflationary bursts will occur earlier if the seed drives a larger Langmuir wave with a shorter $\tau_B$, so that the trapped particles accumulate bounces faster.  We  can increase the Langmuir wave amplitude by increasing the intensity, the duration, or choosing a seed wavelength that produces a higher gain.

\begin{figure}[tbp]
	\includegraphics[width=\linewidth]{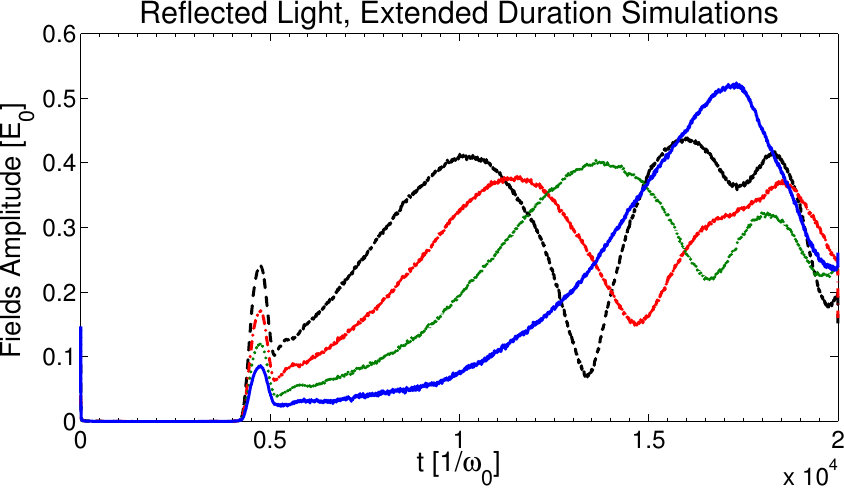}
\caption{The reflected light in extended duration simulations using a Gaussian-like seed pulse with $\lambda_1=1.644\lambda_0$ and various maximum initial intensities.  Shown are $I_{1s}=4\times10^{-3}I_0$ (solid blue), $8\times10^{-3}I_0$ (dotted green), $0.016I_0$ (dash-dotted red), and $0.032I_0$ (dashed black).\label{vs_seed_I}}
\end{figure}

We first vary the initial amplitude of the seed.  Figure \ref{vs_seed_I} shows that the first burst of reflected light has a maximum around $t=17,000\omega_0^{-1}$ when we use a Gaussian-like seed pulse with $I_{1s}=4\times10^{-3}I_0$ and $\lambda_1=1.644\lambda_0$.  As we increase $I_{1s}$, the burst moves earlier, but the difference in amplitude between the burst and final seed amplitude also decreases. As a clear demonstration of the effect of the higher intensity seeds, in Table \ref{table_bounce} we examine the plasma wave amplitudes and particle bounce times for the simulations in Figure \ref{vs_seed_I}.  We examine a selection of particles near the plasma wave phase velocity between $x=350 c/\omega_0$ and $x=450 c/\omega_0$ at $t=5,000\omega_0^{-1}$.  We use the most deeply trapped particle to determine the time it takes to complete one bounce.  To calculate the theoretical bounce time, we filter out all plasma wave modes except $1.4\omega_0/c \leq k \leq 1.5\omega_0/c$ and average the field amplitude over the bounce time along the particle's track.  We then substitute the measured values of the amplitude and wavenumber ($k=1.44\omega_0/c$) into the formula for the bounce time.  The results are in Table \ref{table_bounce}.  The measured bounce times are slightly ($\sim10\%$) longer than the simple expression. We believe this discrepancy is due to the fact that the amplitude of the wave is changing with time and because the calculation is for a parabolic potential well, while the particles are actually trapped in a sinusoidal well.

\begin{table}
\begin{tabular}{l   l   l       l}
$I_{1s}$ &  Avg. EPW Amp.  ~~&  Calc. $\tau_B$  ~~&  Meas. $\tau_B$  \\
$[I_0]$ & $[E_0]$ & $[\omega_0^{-1}]$ & $[\omega_0^{-1}]$ \\
\hline
$4\times10^{-3}$ ~~&  0.020  &  610  &  675\\
$8\times10^{-3}$ &  0.033  &  480  &  550\\
$0.016$ &  0.049  &  390  &  415\\ 
$0.032$ &  0.071  &  320  &  350\\
\hline
\end{tabular}
\caption{The bounce times of deeply trapped electrons measured in the simulations of Figure \ref{vs_seed_I}, along with the bounce times calculated using the average plasma wave field amplitude along the particle's trajectory.\label{table_bounce}}
\end{table}


Besides lowering the kinetic damping rate, another well-known effect of particle trapping is the nonlinear frequency down-shift of the plasma wave.\cite{morales:nonlinear}  As the plasma wave grows, it will shift downward in frequency because it will trap more particles.  According to the frequency matching condition in Eq.\ \ref{w_cond},  the down-shift in the frequency of the plasma wave should be accompanied by an up-shift in the frequency of the scattered light.  We examine this down-shift using a Wigner transform with a Choi filter.\cite{mallat:wavelet}  The Wigner transform takes a function of  time and computes its representation as a function of both frequency and time.  It maps $f(t)\to f(\omega,t)$.

\begin{figure}[tbp]
	\includegraphics[width=\linewidth]{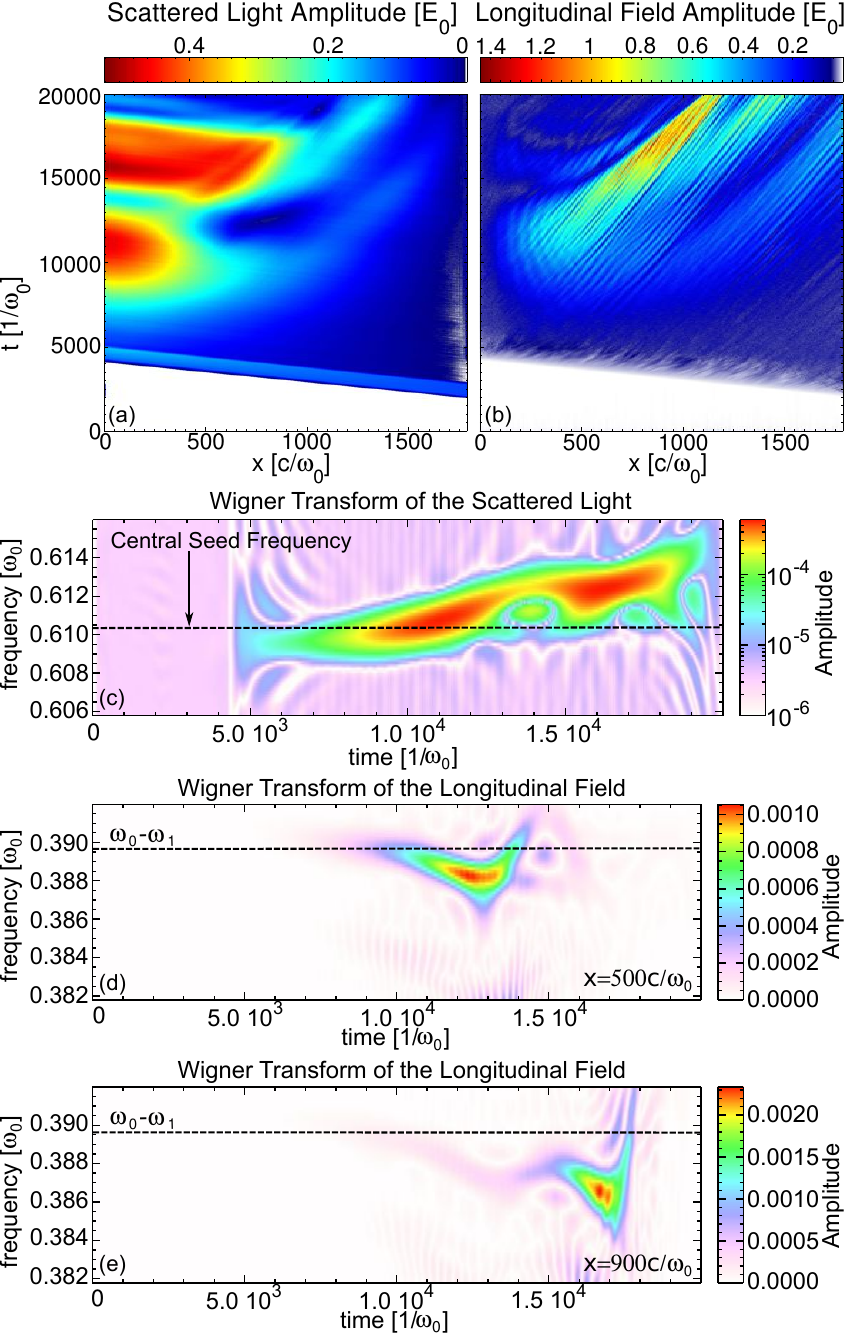}
\caption{The scattered light (a) and plasma wave (b) in a simulation using a flat-top seed pulse with $I_{1s}=8\times10^{-3}I_0$ and $\lambda_1=1.638\lambda_0$.  Below them are the Wigner transforms of the reflected light at $x=0$ (c), the plasma wave at $x=500c/\omega_0$ (d), and the plasma wave at $x=900c/\omega_0$ (e).\label{575nm_wigner}}
\end{figure}

\begin{figure}[tbp]
	\includegraphics[width=\linewidth]{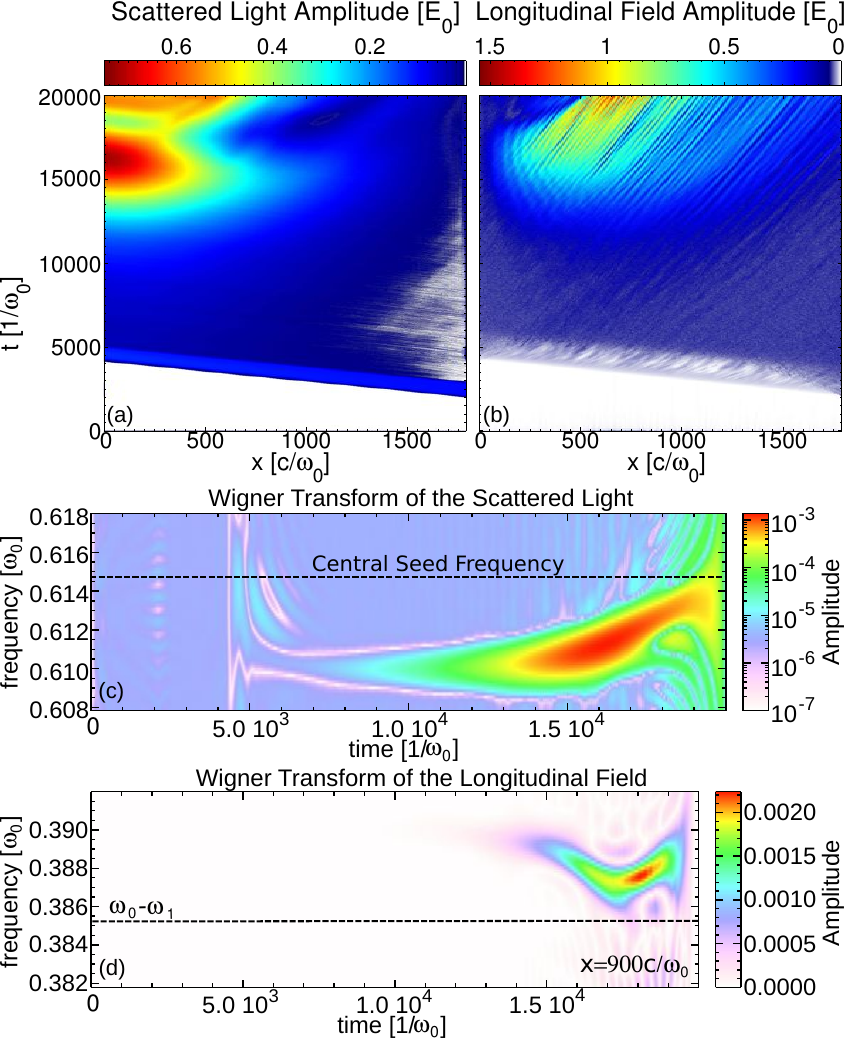}
\caption{The scattered light (a) and plasma wave (b) in a simulation using a flat-top seed pulse with $I_{1s}=8\times10^{-3}I_0$ and $\lambda_1=1.627\lambda_0$.  Below them are the Wigner transforms of the reflected light at $x=0$ (c) and the plasma wave at $x=900c/\omega_0$ (d).\label{571nm_wigner}}
\end{figure}

The Wigner transform results for a run with a flat-top seed with $I_{1s}=8\times10^{-3}I_0$ and $\lambda_1=1.638\lambda_0$ are shown in Figure \ref{575nm_wigner}.  The results for the same run, except using a seed with $\lambda_1=1.627\lambda_0$, are in Figure \ref{571nm_wigner}.   In both cases, the seed appears in the Wigner transform scattered light plots around $t=4500\omega_0^{-1}$, when the seed reaches the left side of the box. The scattered light frequency shifts up while the plasma wave frequency shifts down, as expected due to trapping.

The inflationary bursts begin growing near the frequency with the highest gain regardless of the seed's central frequency, consistent with a harmonic oscillator that is driven off-resonance.  The initial growth is near the central frequency of the seed in the case with $\lambda_1=1.638\lambda_0$, and in the seed's lower-frequency tail in the case with $\lambda_1=1.627\lambda_0$.  This observation suggests that the ponderomotive beating of the seed and the pump simply disturbs the plasma and provides an initial level for growth, but the subsequent growth occurs at the most unstable mode.


There are two inflationary bursts in the scattered light plot in Figure \ref{575nm_wigner}.  According to Ref. \onlinecite{Winjum:Effects}, the separation of the bursts in time is $2\pi/\Delta\omega_{NL}$, where $\Delta\omega_{NL}$ is the nonlinear shift of the scattered light and plasma wave from their non-inflationary resonant frequencies, where peak linear gain occurs.  In our case, the bursts occur at different frequencies.  The first is roughly on-resonance while the second has $\Delta\omega_{NL}  \approx 0.002\omega_0$.  Averaging the two, we have $\Delta\omega_{NL}^{avg}=0.001\omega_0$, or a separation of about $6,000\omega_0^{-1}$, which is in good agreement with the actual separation of the two bursts.  Figures \ref{575nm_wigner} and \ref{571nm_wigner} show the seed wavelength also affects the time required for inflation to set in.  We explore this effect in Sections \ref{measure_inflation} and \ref{continuous}.

\begin{figure}[tbp]
	\includegraphics[width=\linewidth]{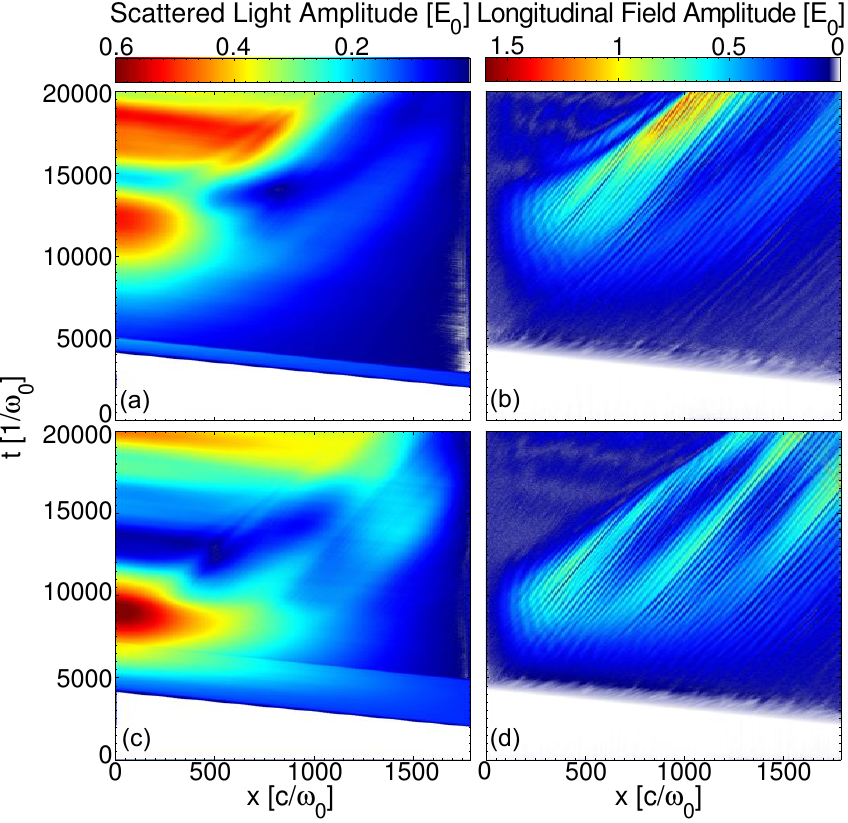}
\caption{Results from simulations with flat-top seed pulses with $I_{1s}=8\times10^{-3}I_0$ and $\lambda_1=1.644\lambda_0$.  The scattered light (left) and plasma wave (right) for simulations using a pulse duriation of $(1000, 3000)\omega_0^{-1}$ are in panels (a-b, c-d), respectively.\label{vspl}}
\end{figure}

Increasing the width of the seed drives the plasma wave longer, increasing its amplitude and causing inflationary bursts to occur earlier.  In Figure \ref{vspl}, we vary the duration of a flat-top pulse while keeping the rise and fall time at $200\omega_0^{-1}$.  The burst of high reflectivity moves earlier and, correspondingly, the plasma wave reaches a high amplitude quicker as we increase the pulse duration from $1,000\omega_0^{-1}$ to $3,000\omega_0^{-1}$.  However, this effect only occurs if we drive near resonance.  If we drive off resonance, then the the large bursts don't occur, and the reflectivity oscillates with a period of $2\pi/\Delta\omega_{NR}$, where $\Delta\omega_{NR}$ is the difference between the seed frequency and the resonant frequency, at which inflationary SRS grows.

\section{Inflation of the Seed}\label{measure_inflation}
\begin{figure}[tbp]
	\includegraphics[width=\linewidth]{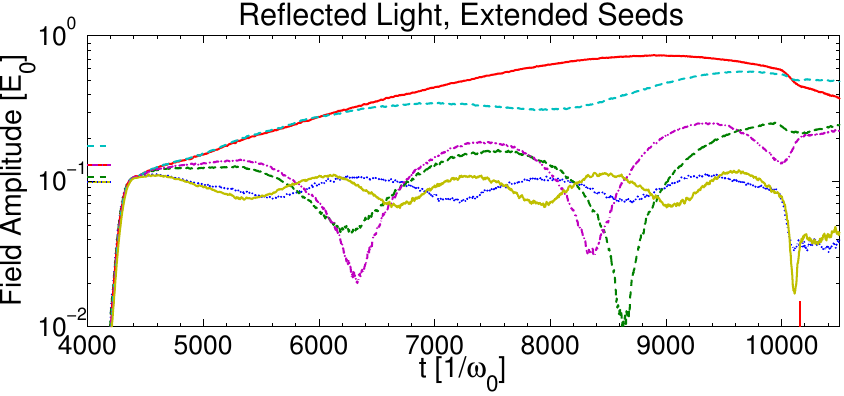}
\caption{The reflected light seen in several simulations using a flat-top seed of duration $6,000\omega_0^{-1}$ with $I_{1s}=8\times10^{-3}I_0$ for various wavelengths.  The two top curves are for seeds with $\lambda_1=1.638\lambda_0$ (solid red) and $1.644\lambda_0$ (dashed cyan).  The four lower curves are for seeds with $\lambda_1=1.627\lambda_0$ (dotted blue), $1.632\lambda_0$ (dashed green), $1.650\lambda_0$ (dash-dotted purple), and $1.658\lambda_0$ (solid yellow).  For comparison, we mark the steady-state linear relativistic PIC values using horizontal dashes on the left side of the plot.  The red vertical dash on the lower right side of the plot indicates approximately when the seeds end.\label{reflect_longpulse}}
\end{figure}

When the seed pulse width becomes comparable to the bounce time, then the seed itself can undergo inflation, illustrating the difference between the linear and inflationary regimes. We perform simulations using a flat-top pulse with a duration of $6,000\omega_0^{-1}$  with $I_{1s}=8\times10^{-3}I_0$ at various wavelengths.  The reflectivity plot in Figure \ref{reflect_longpulse} demonstrates the variation of the scattered light with time in the simulations as compared to the steady-state values from linear theory.  The reflectivities seen in the simulations can significantly exceed the linear theory values, particularly for the seeds with $\lambda_1=1.638\lambda_0$ and $1.644\lambda_0$.  The runs using seeds with $\lambda_1=1.632\lambda_0$ and $1.650\lambda_0$ reach levels above the linear theory values, but dip below linear values several times due to driving off resonance.  In these four cases, inflationary scattering continues after the seed ends around $\omega_0t=10,200$. When we use seeds with $\lambda_1=1.627\lambda_0$ and $1.658\lambda_0$, the reflectivity does not reach significantly above the linear value and drops down when the seed ends, because inflationary scattering does not occur easily when the seed is far from resonance.

We observe oscillations in the reflectivity in all the cases, except when we use the seeds with $\lambda_1=1.638\lambda_0$ and $1.644\lambda_0$.  These oscillations are due to the scattered light seed driving SRS off resonance.\cite{Winjum:Effects}  The ponderomotive beating of the pump and scattered light drives the plasma wave, so if the phase of the beat drive leads the plasma wave by $\pi/2$, the plasma wave no longer grows.  The plasma wave density $n_1 \propto -\frac{\partial}{\partial x}E_2$, while the beat drive $F_p \propto -\frac{\partial}{\partial x}E_0E_1$.  The product, 
\begin{equation}R_p = \left(\frac{\partial}{\partial x}E_2\right)\left(\frac{\partial}{\partial x}E_0E_1\right)\label{Res},\end{equation}
indicates the phase difference between the beat drive and the plasma wave.\cite{winjum:dissertation}  We also note that the change in the pump energy density with time is given by
\begin{equation}
\frac{\partial W_0}{\partial t} = -\frac{e}{8\pi m_e \omega_1 k_2}R_p.
\end{equation}
Therefore, if $R_p$ is positive, the waves are in phase, the beat drive is resonantly driving the plasma wave, and energy is transferred from the pump to the seed and plasma wave.  The inverse applies if $R_p$ is negative.

\begin{figure}[tbp]
	\includegraphics[width=\linewidth]{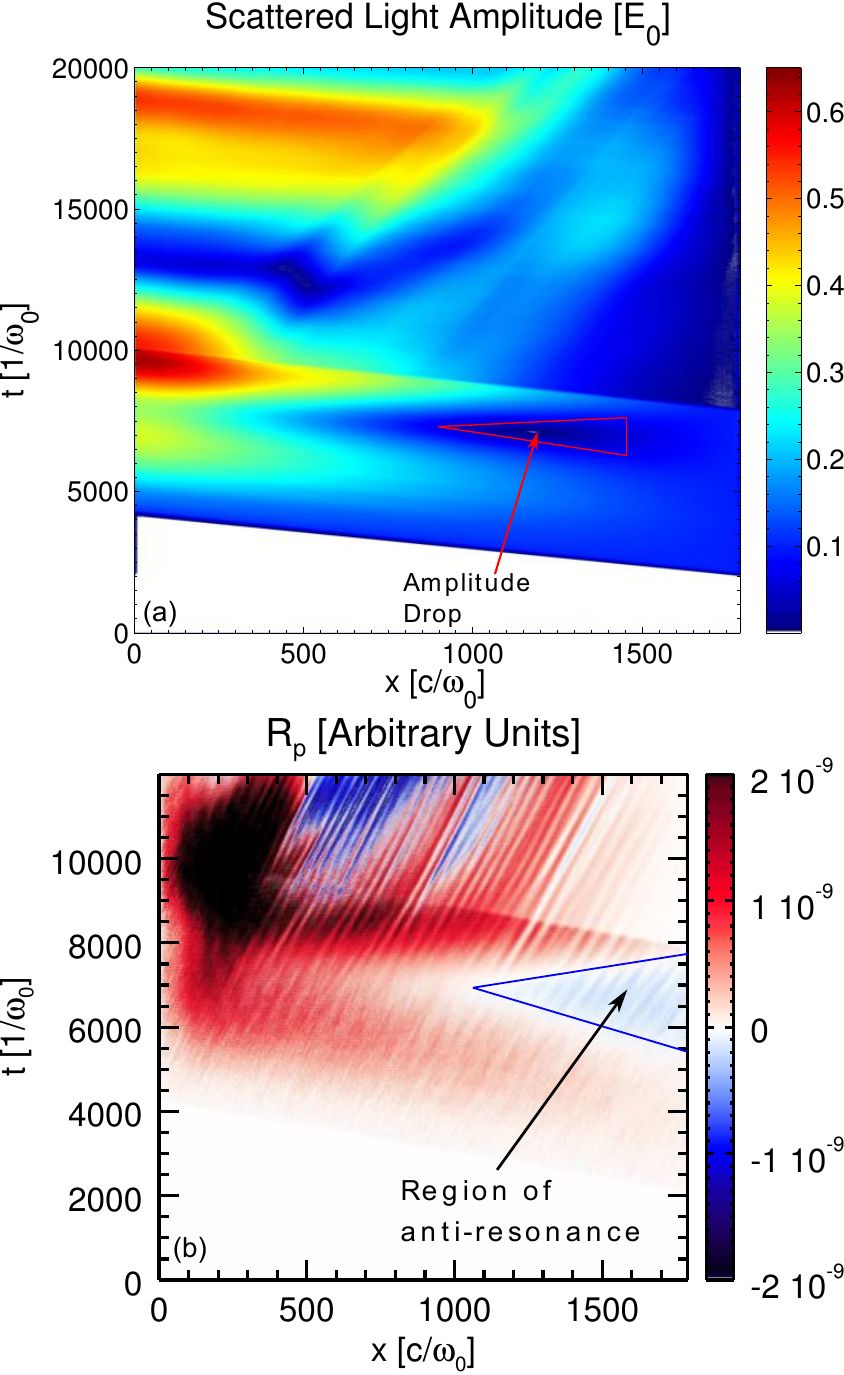}
\caption{The scattered light (a), and the resonance product $R_p$ from Eq. \ref{Res} (b) for the flat-top seed of duration $6,000\omega_0^{-1}$ with $\lambda_1=1.644\lambda_0$.\label{577nm_6000w0t}}
\end{figure}

Figure \ref{577nm_6000w0t} shows the scattered light and the resonance plot for the seed with $\lambda_1=1.644\lambda_0$.  We smooth the result from the resonance diagnostic in $x$ using a 6-point moving average.  Notice the dip (valley)  in the scattered light amplitude around $x=1250c/\omega_0$, $t=7,000\omega_0^{-1}$, and the corresponding negative area on the resonance plot around $x=1500c/\omega_0$.  When this drop occurs, the beat drive and the plasma wave are out of phase, so energy flows from the scattered light wave to the pump.  This shift away from resonance is due to the nonlinear frequency shift of the plasma wave, explained in Section \ref{late_time}.  The non-resonant drive is also responsible for the oscillations we see in the reflectivity plot of Figure \ref{reflect_longpulse}.

%

\begin{figure}[tbp]
	\includegraphics[width=\linewidth]{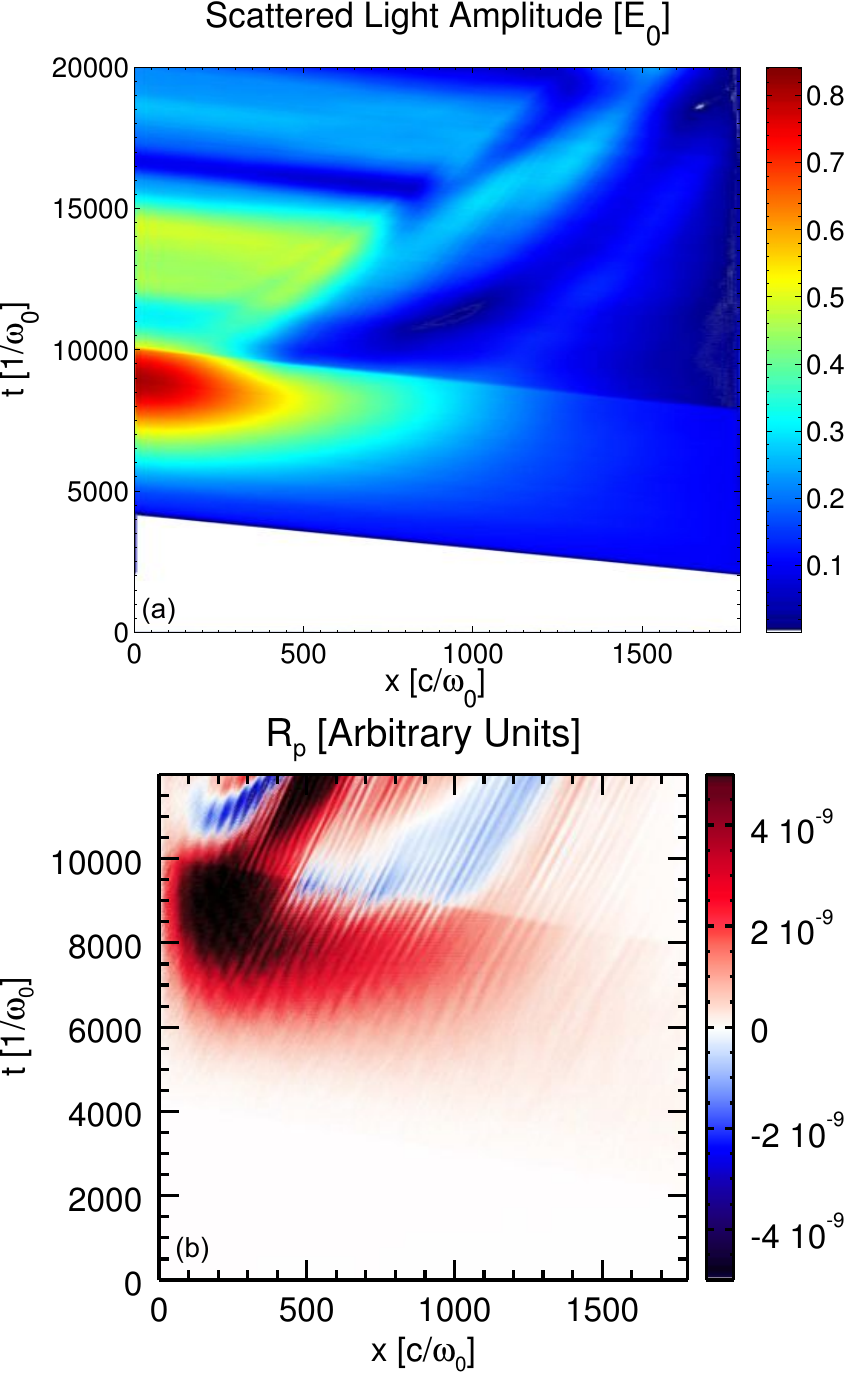}
\caption{The scattered light (a), and the resonance product $R_p$ from Eq. \ref{Res} (b) for the flat-top seed of duration $6,000\omega_0^{-1}$ with $\lambda_1=1.638\lambda_0$.\label{575nm_6000w0t}}
\end{figure}

Figure \ref{575nm_6000w0t} shows the same plots as Figure \ref{577nm_6000w0t}, except using a seed with $\lambda_1=1.638\lambda_0$.  This seed continuously drives the scattered light to its maximum amplitude before dropping off, whereas the one with $\lambda_1=1.644\lambda_0$ drives a lower-amplitude burst of scattered light before it produces a second burst at much higher amplitude.  Notice that the resonance plot in Figure \ref{575nm_6000w0t} shows that the beat wave and plasma wave are in resonance until the peak of the scattered light burst.  This resonant drive leads to a burst of scattered light at higher amplitude than in the $\lambda_1=1.644\lambda_0$ case.

\section{Inflation of Continuous Seeds}\label{continuous}
In this section, we extend the flat-top seed pulse so that it remains on through the end of the simulation.  Based on what we have observed in the Sections \ref{late_time} and \ref{measure_inflation}, we expect the inflationary behavior we see in these simulations to depend on the seed intensity and wavelength.  We know that there must be an intensity threshold below which the seed does not drive inflationary behavior in the simulations (very long bounce times), because we see negligible scattering without a seed.  However, as we saw in Section \ref{measure_inflation}, a seed that is intense enough to drive inflation on resonance may not be intense enough to drive it off resonance.


Furthermore, the use of continuous seeds is related to past work by Winjum et. al. on scattering off of plasma wave packets.\cite{Winjum:Effects}  Light scattering off a wave packet that has undergone a nonlinear frequency shift acts as a seed for BSRS in the unperturbed background plasma.  However, the beat drive frequency is not a natural mode of the background plasma.  This non-resonant beat drive leads to oscillations in the reflected light with a period $2\pi/\Delta\omega_{NL}$, where $\Delta\omega_{NL}$ is the nonlinear frequency shift of the plasma wave, much like the amplitude of a simple harmonic oscillator varies when driven off resonance.  In this section, we examine the effect of resonant and non-resonant drive in more detail by using continuous seeds at different wavelengths and intensities, and examining the results using plots of the reflected light.

\begin{figure}[tbp]
	\includegraphics[width=\linewidth]{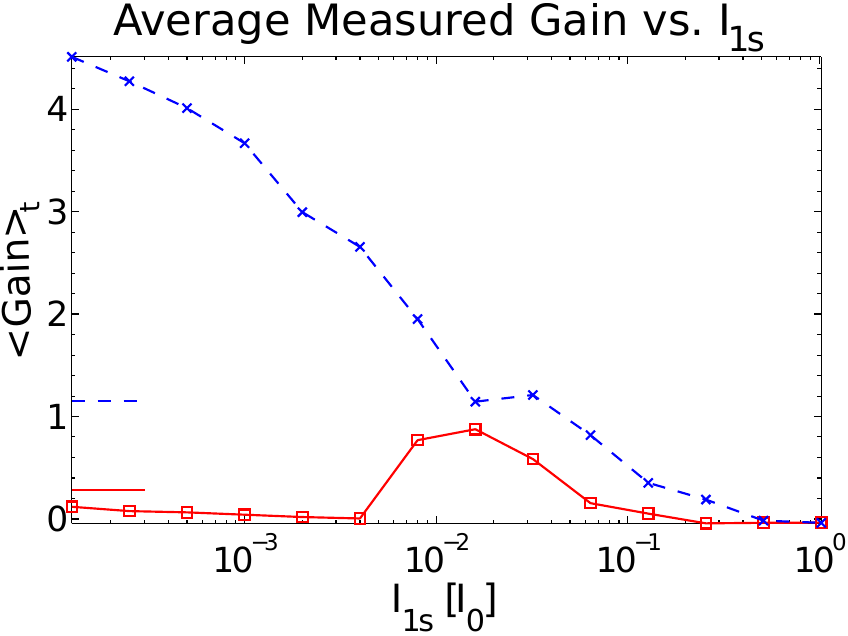}
\caption{The time-average measured gain seen in simulations as we vary the continuous seed intensity using seeds with $\lambda_1=1.644\lambda_0$ (dashed blue) and $1.658\lambda_0$ (solid red).  For comparison, we use horizontal dashes on the left side of the plot to mark the steady-state gain from linear relativistic PIC theory.\label{continuous_gainvI1}}
\end{figure}

Figure \ref{continuous_gainvI1} shows the time-average measured gain for various seed intensities with $\lambda_1=1.644\lambda_0$ and $1.658\lambda_0$.  When we use a seed with $\lambda_1=1.644\lambda_0$, an intensity of $1.25\times10^{-4}I_0$ (smallest value shown) is enough to cause inflation.  However, when we use a seed with $\lambda_1=1.658\lambda_0$, we do not see inflation until the seed intensity reaches $8\times10^{-3}I_0$.  The measured gain at both wavelengths decreases with seed intensity due to pump depletion once inflation sets in.

In order to understand the results with a $1.658\lambda_0$ seed, we first discuss trapping effects on the SRS gain spectrum via Eq. \ref{g0}.  Trapping nonlinearity leads to a reduction in $\chi_i$ and thus Landau damping, as well as a down-shift in the natural plasma wave frequency. This latter effect decreases the resonant scattered wavelength, where $1+\chi_r=0$.  In electrostatic simulations with an external driver, Fahlen showed behavior consistent with this picture.\cite{fahlen:dissertation} For $k_2\lambda_{De}\sim0.3$ and a driver frequency above the natural frequency, the plasma response is even smaller than the linear response. However, when the driver is below the natural frequency, a larger response is obtained.

\begin{figure}[tbp]
	\includegraphics[width=\linewidth]{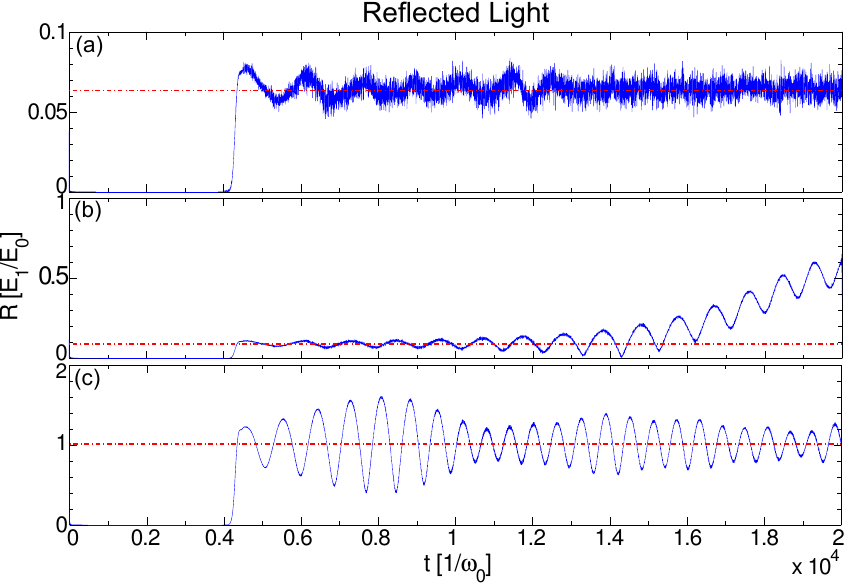}
	\caption{The reflected light in simulations using a continuous seed with $\lambda_1=1.658\lambda_0$ (solid red in Fig. \ref{continuous_gainvI1}).  $I_{1s}/I_0=4\times10^{-3}$ (a), $8\times10^{-3}$ (b), and $1.024I_0$ (c).  The horizontal dash-dotted line indicates the amplitude of the seed (unamplified) reflected light.\label{582nm_refl}}
\end{figure}

Trapping, therefore, makes the gain spectrum narrower and peaked at a smaller wavelength. The gain increases for wavelengths near resonance. However, for wavelengths far from resonance, the gain does \textit{not} increase, and, in fact, vanishes as $\chi_i \rightarrow 0$.  We can see this effect clearly by examining Eq. \ref{g0}.  Since the seed is far from resonance, $1+\chi_r \gg \chi_i$, so $g_0 \propto \chi_i/(1+\chi_r)^2$.  This vanishing gain is clear in Figure \ref{582nm_refl}a, where the gain approaches zero in steady state for the continuous $1.658\lambda_0$ seed with $I_{1s} / I_0 = 4\times10^{-3}$.  Thus, inflation is not possible at non-resonant wavelengths.  For hot, low-density plasmas, there is no resonant wavelength, and all phase-matched plasma waves satisfy the loss of resonance condition $k_2 \lambda_{De} > 0.53$.\cite{rose:self-consistent}  Inflation cannot occur in such a plasma at any scattered wavelength.

The red curve in Figure \ref{continuous_gainvI1} is for a seed wavelength that is non-resonant and larger than the linear resonance.  The reduction of $\chi_i$ due to trapping cannot lead to inflation at this wavelength, and the nonlinear frequency shift will move the resonance farther away. Both effects conspire to reduce the SRS gain below its linear value, which is what we observe for the lowest seed intensities.  The number of bounce orbits completed by resonant electrons,\cite{strozzi:kinetic, strozzi:characterizing} based on the plasma wave amplitude computed from linear theory, is $>4$.  It is thus consistent for trapping nonlinearity to occur and reduce the SRS gain.  The increase in gain for $I_{1s} / I_0 = 8\times10^{-3}$, as shown in Figure \ref{582nm_refl}b, first develops at linear resonance, $1.644\lambda_0$, \textit{not} at the seed value of $1.658\lambda_0$, then shifts down in wavelength with time to finish around $1.638\lambda_0$.  This progression is similar to the one in Figure \ref{571nm_wigner}.  We performed a similar run using a flat-top seed of duration $1,000\omega_0^{-1}$ with a central wavelength of $1.658\lambda_0$ and $I_{1s} / I_0 = 4\times10^{-3}$ and observed inflation similar to that in Figure \ref{582nm_refl}b, but without the oscillations, which raises the possibility that continuous seeds can suppress inflation.

%
%
There are oscillations in Figure \ref{582nm_refl}b, which occur because the seed is driving SRS off resonance.  As we discussed earlier, this non-resonant drive leads to oscillations in the reflected light with a period of $2\pi/\Delta\omega_{NR}$, where $\Delta\omega_{NR}$ is the difference between the seed frequency and the resonant frequency, at which inflationary SRS grows. Equivalently, the $\Delta\omega_{NR}$ is the difference between the seeded beat drive frequency and the frequency of the plasma wave packet.  When we use a seed with $I_{1s}/I_0=1.024$, the oscillations are more prominent and faster, and we see no amplification, as seen in Figure \ref{582nm_refl}c.  The increased oscillation frequency is due to the higher amplitude plasma wave undergoing a greater frequency shift.  We also see a beat wave pattern covering many oscillations, which is caused by the nonlinear frequency shift of the plasma wave packet, as described earlier in this section.

\begin{figure}[tbp]
	\includegraphics[width=\linewidth]{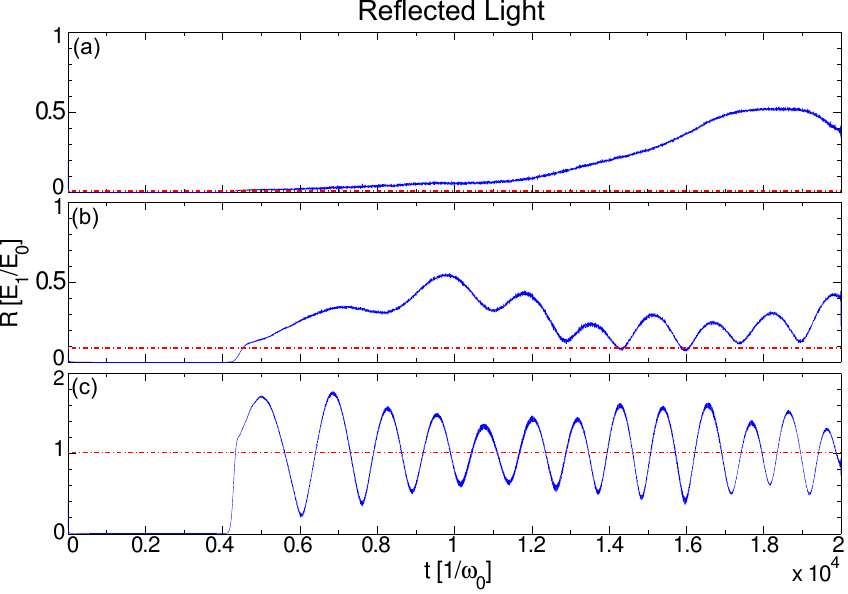}
	\caption{The reflected light in simulations using a continuous seed with $\lambda_1=1.644\lambda_0$ (dashed blue in Fig. \ref{continuous_gainvI1}).  $I_{1s}/I_0=1.25\times10^{-4}$ (a), $8\times10^{-3}$ (b), and $1.024$ (c).  The horizontal dash-dotted line indicates the amplitude of the seed (unamplified) reflected light.\label{577nm_refl}}
\end{figure}

Figure \ref{577nm_refl} shows the reflected light in simulations using a (low-, moderate-, high-) intensity seed with $I_{1s}/I_0=(1.25\times10^{-4},8\times10^{-3}, 1.024)$, but $\lambda_1=1.644\lambda_0$.  When we use the low-intensity seed, we see the reflected light increase monotonically until it saturates near the end of the simulation.  Unlike when we used $\lambda_1=1.658\lambda_0$, this seed is near resonance, so $1+\chi_r << \chi_i$, and $g_0 \propto 1/\chi_i$.  Therefore, particle trapping increases the gain, as explained earlier in the paper.  Oscillations begin to appear again in the simulation with the moderate-intensity seed, which indicates that the seed is slightly off resonance, as we expect to occur as the plasma wave undergoes a nonlinear frequency shift.  These oscillations once again become more prominent when we use the high-intensity seed, and we see a beat wave pattern appear again.

\section{conclusions and future research}\label{conclusion}
Using 1D OSIRIS PIC simulations, we have studied BSRS of a well-defined seed pulse with variable shape, intensity, and wavelength. We found that backward Raman amplification of a seed scattered light pulse can remain in the strongly damped convective regime.  That is, for a sufficiently weak seed, kinetic inflation does not occur.  Peak seed amplification occurs near the peak of the linear gain spectrum when we take into account special relativity and PIC effects such as finite particle size, finite-difference operators, and field smoothing.  Detailed comparisons with linear coupled-mode predictions for envelope dynamics show excellent agreement.  If the seed pulse is intense enough, the driven plasma wave traps particles, thereby lowering the Landau damping after a time on the order of a bounce period.  The plasma wave continues to scatter light and grows after the seed leaves the system, leading to kinetic inflation later in the simulation.  When we extend the seed in time, kinetic inflation occurs while the seed is still present, and we measure dynamic seed amplification, which significantly exceeds the linear gain rate at times, and can also turn negative at other times (i.e., it transfers energy into the pump). When we use a continuous seed, we find that the onset of inflation depends strongly on the seed wavelength.  When the continuous seed wavelength is near the peak of the gain curve, we see kinetic inflation occur with very low seed amplitudes, while higher seed amplitudes are necessary to drive kinetic inflation using a non-resonant seed.

Connecting kinetic inflation with experiments involving lasers with many hot spots or speckles is still at an early stage.  Past research has demonstrated that hot electrons, beam acoustic modes, and side-scattered light can couple hot-spots transversely in 2D simulations, leading to higher BSRS from each hot spot than one would see without coupling.\cite{albright:particle-in-cell,Yin:Self-Organized}    Our results suggest that we can expect to see inflationary scattering from the lower-intensity parts of the beam interacting with flattened (non-Maxwellian) distributions or plasma waves generated in more intense parts, or increased scattered light from a few speckles triggering a ``chain reaction'' of downstream inflation.  If this chain reaction occurs, most of the Raman in underdense laser-produced plasmas, such as ICF targets, will be inflationary.  PIC codes can be used to model hundreds of speckles, but to simulate large volumes across a hohlraum effectively requires the use of envelope codes (such as pF3D), for which reduced models of kinetic nonlinearity are being pursued by several groups.\cite{benisti:nonlinear, Yampolsky:Simplified, Lindberg:Self-consistent}

An important factor we have not explored is de-trapping mechanisms.  All of our simulations in this paper are 1D and ``collisionless," which means that plasma waves easily trap electrons, and trapped electrons cannot leave sideways or be scattered out of the plasma wave.  Since particle trapping is necessary for inflationary scattering, any de-trapping mechanisms make it less likely to occur and impose a threshold amplitude for inflation (the threshold in the present work is set by the finite amplitude and duration of the seed).  In particular, collisions can kick electrons out of a plasma wave's potential well, and electrons can traverse the plasma wave in less than one bounce period in higher dimensions.\cite{fahlen:transverse, strozzi:characterizing}  Future research in this area should explore the effect of these de-trapping processes on inflation, and validate reduced descriptions.

\begin{acknowledgments}
This work performed under the auspices of the U.S. Department of Energy by Lawrence Livermore National Laboratory under Contract DE-AC52-07NA27344 and by the University of California, Los Angeles under Grants FG52-09NA29552 and DE-NA0001833.  This work was funded in part by the Laboratory Directed Research and Development Program at LLNL under project tracking code 08-ERD-017.  Work by I. N. Ellis was supported in part by the Lawrence Scholar Program at LLNL.
\end{acknowledgments}

\bibliography{SRSPoP}

\providecommand{\noopsort}[1]{}\providecommand{\singleletter}[1]{#1}%
\begin{thebibliography}{50}%
\makeatletter
\providecommand \@ifxundefined [1]{%
 \@ifx{#1\undefined}
}%
\providecommand \@ifnum [1]{%
 \ifnum #1\expandafter \@firstoftwo
 \else \expandafter \@secondoftwo
 \fi
}%
\providecommand \@ifx [1]{%
 \ifx #1\expandafter \@firstoftwo
 \else \expandafter \@secondoftwo
 \fi
}%
\providecommand \natexlab [1]{#1}%
\providecommand \enquote  [1]{``#1''}%
\providecommand \bibnamefont  [1]{#1}%
\providecommand \bibfnamefont [1]{#1}%
\providecommand \citenamefont [1]{#1}%
\providecommand \href@noop [0]{\@secondoftwo}%
\providecommand \href [0]{\begingroup \@sanitize@url \@href}%
\providecommand \@href[1]{\@@startlink{#1}\@@href}%
\providecommand \@@href[1]{\endgroup#1\@@endlink}%
\providecommand \@sanitize@url [0]{\catcode `\\12\catcode `\$12\catcode
  `\&12\catcode `\#12\catcode `\^12\catcode `\_12\catcode `\%12\relax}%
\providecommand \@@startlink[1]{}%
\providecommand \@@endlink[0]{}%
\providecommand \url  [0]{\begingroup\@sanitize@url \@url }%
\providecommand \@url [1]{\endgroup\@href {#1}{\urlprefix }}%
\providecommand \urlprefix  [0]{URL }%
\providecommand \Eprint [0]{\href }%
\providecommand \doibase [0]{http://dx.doi.org/}%
\providecommand \selectlanguage [0]{\@gobble}%
\providecommand \bibinfo  [0]{\@secondoftwo}%
\providecommand \bibfield  [0]{\@secondoftwo}%
\providecommand \translation [1]{[#1]}%
\providecommand \BibitemOpen [0]{}%
\providecommand \bibitemStop [0]{}%
\providecommand \bibitemNoStop [0]{.\EOS\space}%
\providecommand \EOS [0]{\spacefactor3000\relax}%
\providecommand \BibitemShut  [1]{\csname bibitem#1\endcsname}%
\let\auto@bib@innerbib\@empty
\bibitem [{\citenamefont {Drake}\ \emph {et~al.}(1974)\citenamefont {Drake},
  \citenamefont {Kaw}, \citenamefont {Lee}, \citenamefont {Schmid},
  \citenamefont {Liu},\ and\ \citenamefont {Rosenbluth}}]{drake:parametric}%
  \BibitemOpen
  \bibfield  {author} {\bibinfo {author} {\bibfnamefont {J.~F.}\ \bibnamefont
  {Drake}}, \bibinfo {author} {\bibfnamefont {P.~K.}\ \bibnamefont {Kaw}},
  \bibinfo {author} {\bibfnamefont {Y.~C.}\ \bibnamefont {Lee}}, \bibinfo
  {author} {\bibfnamefont {G.}~\bibnamefont {Schmid}}, \bibinfo {author}
  {\bibfnamefont {C.~S.}\ \bibnamefont {Liu}}, \ and\ \bibinfo {author}
  {\bibfnamefont {M.~N.}\ \bibnamefont {Rosenbluth}},\ }\href {\doibase
  10.1063/1.1694789} {\bibfield  {journal} {\bibinfo  {journal} {Physics of
  Fluids}\ }\textbf {\bibinfo {volume} {17}},\ \bibinfo {pages} {778} (\bibinfo
  {year} {1974})}\BibitemShut {NoStop}%
\bibitem [{\citenamefont {Forslund}, \citenamefont {Kindel},\ and\
  \citenamefont {Lindman}(1975{\natexlab{a}})}]{forslund:theory}%
  \BibitemOpen
  \bibfield  {author} {\bibinfo {author} {\bibfnamefont {D.~W.}\ \bibnamefont
  {Forslund}}, \bibinfo {author} {\bibfnamefont {J.~M.}\ \bibnamefont
  {Kindel}}, \ and\ \bibinfo {author} {\bibfnamefont {E.~L.}\ \bibnamefont
  {Lindman}},\ }\href {\doibase 10.1063/1.861248} {\bibfield  {journal}
  {\bibinfo  {journal} {Physics of Fluids}\ }\textbf {\bibinfo {volume} {18}},\
  \bibinfo {pages} {1002} (\bibinfo {year} {1975}{\natexlab{a}})}\BibitemShut
  {NoStop}%
\bibitem [{\citenamefont {Forslund}, \citenamefont {Kindel},\ and\
  \citenamefont {Lindman}(1975{\natexlab{b}})}]{forslund:simulation}%
  \BibitemOpen
  \bibfield  {author} {\bibinfo {author} {\bibfnamefont {D.~W.}\ \bibnamefont
  {Forslund}}, \bibinfo {author} {\bibfnamefont {J.~M.}\ \bibnamefont
  {Kindel}}, \ and\ \bibinfo {author} {\bibfnamefont {E.~L.}\ \bibnamefont
  {Lindman}},\ }\href {\doibase 10.1063/1.861249} {\bibfield  {journal}
  {\bibinfo  {journal} {Physics of Fluids}\ }\textbf {\bibinfo {volume} {18}},\
  \bibinfo {pages} {1017} (\bibinfo {year} {1975}{\natexlab{b}})},\ \bibinfo
  {note} {{And references therein}}\BibitemShut {NoStop}%
\bibitem [{\citenamefont {Atzeni}\ and\ \citenamefont {Meyer-ter
  Vehn}(2009)}]{atzeni:physics}%
  \BibitemOpen
  \bibfield  {author} {\bibinfo {author} {\bibfnamefont {S.}~\bibnamefont
  {Atzeni}}\ and\ \bibinfo {author} {\bibfnamefont {J.}~\bibnamefont {Meyer-ter
  Vehn}},\ }\href {http://books.google.com/books?id=2EMgcAAACAAJ} {\emph
  {\bibinfo {title} {{The Physics of Inertial Fusion: Beam Plasma Interaction,
  Hydrodynamics, Hot Dense Matter}}}},\ The International Series of Monographs
  on Physics\ (\bibinfo  {publisher} {Oxford University Press, USA},\ \bibinfo
  {year} {2009})\BibitemShut {NoStop}%
\bibitem [{\citenamefont {Lindl}\ \emph {et~al.}(2004)\citenamefont {Lindl},
  \citenamefont {Amendt}, \citenamefont {Berger}, \citenamefont {Glendinning},
  \citenamefont {Glenzer}, \citenamefont {Haan}, \citenamefont {Kauffman},
  \citenamefont {Landen},\ and\ \citenamefont {Suter}}]{lindl:physics}%
  \BibitemOpen
  \bibfield  {author} {\bibinfo {author} {\bibfnamefont {J.~D.}\ \bibnamefont
  {Lindl}}, \bibinfo {author} {\bibfnamefont {P.}~\bibnamefont {Amendt}},
  \bibinfo {author} {\bibfnamefont {R.~L.}\ \bibnamefont {Berger}}, \bibinfo
  {author} {\bibfnamefont {S.~G.}\ \bibnamefont {Glendinning}}, \bibinfo
  {author} {\bibfnamefont {S.~H.}\ \bibnamefont {Glenzer}}, \bibinfo {author}
  {\bibfnamefont {S.~W.}\ \bibnamefont {Haan}}, \bibinfo {author}
  {\bibfnamefont {R.~L.}\ \bibnamefont {Kauffman}}, \bibinfo {author}
  {\bibfnamefont {O.~L.}\ \bibnamefont {Landen}}, \ and\ \bibinfo {author}
  {\bibfnamefont {L.~J.}\ \bibnamefont {Suter}},\ }\href {\doibase
  10.1063/1.1578638} {\bibfield  {journal} {\bibinfo  {journal} {Physics of
  Plasmas}\ }\textbf {\bibinfo {volume} {11}},\ \bibinfo {pages} {339}
  (\bibinfo {year} {2004})}\BibitemShut {NoStop}%
\bibitem [{\citenamefont {Koch}\ and\ \citenamefont
  {Albritton}(1975)}]{koch:nonlinear}%
  \BibitemOpen
  \bibfield  {author} {\bibinfo {author} {\bibfnamefont {P.}~\bibnamefont
  {Koch}}\ and\ \bibinfo {author} {\bibfnamefont {J.}~\bibnamefont
  {Albritton}},\ }\href {\doibase 10.1103/PhysRevLett.34.1616} {\bibfield
  {journal} {\bibinfo  {journal} {Phys. Rev. Lett.}\ }\textbf {\bibinfo
  {volume} {34}},\ \bibinfo {pages} {1616} (\bibinfo {year}
  {1975})}\BibitemShut {NoStop}%
\bibitem [{\citenamefont {Estabrook}, \citenamefont {Kruer},\ and\
  \citenamefont {Haines}(1989)}]{estabrook:nonlinear}%
  \BibitemOpen
  \bibfield  {author} {\bibinfo {author} {\bibfnamefont {K.}~\bibnamefont
  {Estabrook}}, \bibinfo {author} {\bibfnamefont {W.~L.}\ \bibnamefont
  {Kruer}}, \ and\ \bibinfo {author} {\bibfnamefont {M.~G.}\ \bibnamefont
  {Haines}},\ }\href {\doibase 10.1063/1.858952} {\bibfield  {journal}
  {\bibinfo  {journal} {Physics of Fluids B: Plasma Physics}\ }\textbf
  {\bibinfo {volume} {1}},\ \bibinfo {pages} {1282} (\bibinfo {year}
  {1989})}\BibitemShut {NoStop}%
\bibitem [{\citenamefont {Estabrook}\ and\ \citenamefont
  {Kruer}(1983)}]{Estabrook:Theory}%
  \BibitemOpen
  \bibfield  {author} {\bibinfo {author} {\bibfnamefont {K.}~\bibnamefont
  {Estabrook}}\ and\ \bibinfo {author} {\bibfnamefont {W.~L.}\ \bibnamefont
  {Kruer}},\ }\href {\doibase 10.1063/1.864336} {\bibfield  {journal} {\bibinfo
   {journal} {Physics of Fluids}\ }\textbf {\bibinfo {volume} {26}},\ \bibinfo
  {pages} {1892} (\bibinfo {year} {1983})}\BibitemShut {NoStop}%
\bibitem [{\citenamefont {Kruer}\ \emph {et~al.}(1980)\citenamefont {Kruer},
  \citenamefont {Estabrook}, \citenamefont {Lasinski},\ and\ \citenamefont
  {Langdon}}]{kruer:raman}%
  \BibitemOpen
  \bibfield  {author} {\bibinfo {author} {\bibfnamefont {W.~L.}\ \bibnamefont
  {Kruer}}, \bibinfo {author} {\bibfnamefont {K.}~\bibnamefont {Estabrook}},
  \bibinfo {author} {\bibfnamefont {B.~F.}\ \bibnamefont {Lasinski}}, \ and\
  \bibinfo {author} {\bibfnamefont {A.~B.}\ \bibnamefont {Langdon}},\ }\href
  {\doibase 10.1063/1.863145} {\bibfield  {journal} {\bibinfo  {journal}
  {Physics of Fluids}\ }\textbf {\bibinfo {volume} {23}},\ \bibinfo {pages}
  {1326} (\bibinfo {year} {1980})}\BibitemShut {NoStop}%
\bibitem [{\citenamefont {Montgomery}\ \emph {et~al.}(2002)\citenamefont
  {Montgomery}, \citenamefont {Cobble}, \citenamefont {Fernandez},
  \citenamefont {Focia}, \citenamefont {Johnson}, \citenamefont
  {Renard-LeGalloudec}, \citenamefont {Rose},\ and\ \citenamefont
  {Russell}}]{montgomery:recent}%
  \BibitemOpen
  \bibfield  {author} {\bibinfo {author} {\bibfnamefont {D.~S.}\ \bibnamefont
  {Montgomery}}, \bibinfo {author} {\bibfnamefont {J.~A.}\ \bibnamefont
  {Cobble}}, \bibinfo {author} {\bibfnamefont {J.~C.}\ \bibnamefont
  {Fernandez}}, \bibinfo {author} {\bibfnamefont {R.~J.}\ \bibnamefont
  {Focia}}, \bibinfo {author} {\bibfnamefont {R.~P.}\ \bibnamefont {Johnson}},
  \bibinfo {author} {\bibfnamefont {N.}~\bibnamefont {Renard-LeGalloudec}},
  \bibinfo {author} {\bibfnamefont {H.~A.}\ \bibnamefont {Rose}}, \ and\
  \bibinfo {author} {\bibfnamefont {D.~A.}\ \bibnamefont {Russell}},\ }\href
  {\doibase 10.1063/1.1468857} {\bibfield  {journal} {\bibinfo  {journal}
  {Physics of Plasmas}\ }\textbf {\bibinfo {volume} {9}},\ \bibinfo {pages}
  {2311} (\bibinfo {year} {2002})}\BibitemShut {NoStop}%
\bibitem [{\citenamefont {Froula}\ \emph {et~al.}(2009)\citenamefont {Froula},
  \citenamefont {Divol}, \citenamefont {London}, \citenamefont {Berger},
  \citenamefont {D\"oppner}, \citenamefont {Meezan}, \citenamefont {Ross},
  \citenamefont {Suter}, \citenamefont {Sorce},\ and\ \citenamefont
  {Glenzer}}]{froula:observation}%
  \BibitemOpen
  \bibfield  {author} {\bibinfo {author} {\bibfnamefont {D.~H.}\ \bibnamefont
  {Froula}}, \bibinfo {author} {\bibfnamefont {L.}~\bibnamefont {Divol}},
  \bibinfo {author} {\bibfnamefont {R.~A.}\ \bibnamefont {London}}, \bibinfo
  {author} {\bibfnamefont {R.~L.}\ \bibnamefont {Berger}}, \bibinfo {author}
  {\bibfnamefont {T.}~\bibnamefont {D\"oppner}}, \bibinfo {author}
  {\bibfnamefont {N.~B.}\ \bibnamefont {Meezan}}, \bibinfo {author}
  {\bibfnamefont {J.~S.}\ \bibnamefont {Ross}}, \bibinfo {author}
  {\bibfnamefont {L.~J.}\ \bibnamefont {Suter}}, \bibinfo {author}
  {\bibfnamefont {C.}~\bibnamefont {Sorce}}, \ and\ \bibinfo {author}
  {\bibfnamefont {S.~H.}\ \bibnamefont {Glenzer}},\ }\href {\doibase
  10.1103/PhysRevLett.103.045006} {\bibfield  {journal} {\bibinfo  {journal}
  {Phys. Rev. Lett.}\ }\textbf {\bibinfo {volume} {103}},\ \bibinfo {pages}
  {045006} (\bibinfo {year} {2009})}\BibitemShut {NoStop}%
\bibitem [{\citenamefont {Vu}, \citenamefont {DuBois},\ and\ \citenamefont
  {Bezzerides}(2001)}]{vu:transient}%
  \BibitemOpen
  \bibfield  {author} {\bibinfo {author} {\bibfnamefont {H.~X.}\ \bibnamefont
  {Vu}}, \bibinfo {author} {\bibfnamefont {D.~F.}\ \bibnamefont {DuBois}}, \
  and\ \bibinfo {author} {\bibfnamefont {B.}~\bibnamefont {Bezzerides}},\
  }\href {\doibase 10.1103/PhysRevLett.86.4306} {\bibfield  {journal} {\bibinfo
   {journal} {Phys. Rev. Lett.}\ }\textbf {\bibinfo {volume} {86}},\ \bibinfo
  {pages} {4306} (\bibinfo {year} {2001})}\BibitemShut {NoStop}%
\bibitem [{\citenamefont {Vu}, \citenamefont {DuBois},\ and\ \citenamefont
  {Bezzerides}(2002)}]{vu:kinetic}%
  \BibitemOpen
  \bibfield  {author} {\bibinfo {author} {\bibfnamefont {H.~X.}\ \bibnamefont
  {Vu}}, \bibinfo {author} {\bibfnamefont {D.~F.}\ \bibnamefont {DuBois}}, \
  and\ \bibinfo {author} {\bibfnamefont {B.}~\bibnamefont {Bezzerides}},\
  }\href {\doibase 10.1063/1.1471235} {\bibfield  {journal} {\bibinfo
  {journal} {Physics of Plasmas}\ }\textbf {\bibinfo {volume} {9}},\ \bibinfo
  {pages} {1745} (\bibinfo {year} {2002})}\BibitemShut {NoStop}%
\bibitem [{\citenamefont {Vu}, \citenamefont {DuBois},\ and\ \citenamefont
  {Bezzerides}(2007)}]{vu:inflation}%
  \BibitemOpen
  \bibfield  {author} {\bibinfo {author} {\bibfnamefont {H.~X.}\ \bibnamefont
  {Vu}}, \bibinfo {author} {\bibfnamefont {D.~F.}\ \bibnamefont {DuBois}}, \
  and\ \bibinfo {author} {\bibfnamefont {B.}~\bibnamefont {Bezzerides}},\
  }\href {\doibase 10.1063/1.2426918} {\bibfield  {journal} {\bibinfo
  {journal} {Physics of Plasmas}\ }\textbf {\bibinfo {volume} {14}},\ \bibinfo
  {eid} {012702} (\bibinfo {year} {2007})}\BibitemShut {NoStop}%
\bibitem [{\citenamefont {Strozzi}\ \emph {et~al.}(2007)\citenamefont
  {Strozzi}, \citenamefont {Williams}, \citenamefont {Langdon},\ and\
  \citenamefont {Bers}}]{strozzi:kinetic}%
  \BibitemOpen
  \bibfield  {author} {\bibinfo {author} {\bibfnamefont {D.~J.}\ \bibnamefont
  {Strozzi}}, \bibinfo {author} {\bibfnamefont {E.~A.}\ \bibnamefont
  {Williams}}, \bibinfo {author} {\bibfnamefont {A.~B.}\ \bibnamefont
  {Langdon}}, \ and\ \bibinfo {author} {\bibfnamefont {A.}~\bibnamefont
  {Bers}},\ }\href {\doibase 10.1063/1.2431161} {\bibfield  {journal} {\bibinfo
   {journal} {Physics of Plasmas}\ }\textbf {\bibinfo {volume} {14}},\ \bibinfo
  {eid} {013104} (\bibinfo {year} {2007})},\ \Eprint
  {http://arxiv.org/abs/physics/0610029} {arXiv:physics/0610029
  [physics.plasm-ph]} \BibitemShut {NoStop}%
\bibitem [{\citenamefont {O'Neil}(1965)}]{oneil:collisionless}%
  \BibitemOpen
  \bibfield  {author} {\bibinfo {author} {\bibfnamefont {T.}~\bibnamefont
  {O'Neil}},\ }\href {\doibase 10.1063/1.1761193} {\bibfield  {journal}
  {\bibinfo  {journal} {Physics of Fluids}\ }\textbf {\bibinfo {volume} {8}},\
  \bibinfo {pages} {2255} (\bibinfo {year} {1965})}\BibitemShut {NoStop}%
\bibitem [{\citenamefont {Morales}\ and\ \citenamefont
  {O'Neil}(1972)}]{morales:nonlinear}%
  \BibitemOpen
  \bibfield  {author} {\bibinfo {author} {\bibfnamefont {G.~J.}\ \bibnamefont
  {Morales}}\ and\ \bibinfo {author} {\bibfnamefont {T.~M.}\ \bibnamefont
  {O'Neil}},\ }\href {\doibase 10.1103/PhysRevLett.28.417} {\bibfield
  {journal} {\bibinfo  {journal} {Phys. Rev. Lett.}\ }\textbf {\bibinfo
  {volume} {28}},\ \bibinfo {pages} {417} (\bibinfo {year} {1972})}\BibitemShut
  {NoStop}%
\bibitem [{\citenamefont {Winjum}(2010)}]{winjum:dissertation}%
  \BibitemOpen
  \bibfield  {author} {\bibinfo {author} {\bibfnamefont {B.~J.}\ \bibnamefont
  {Winjum}},\ }\emph {\bibinfo {title} {Particle-In-Cell Simulations of
  Stimulated Raman Scattering for Parameters Relevant to Inertial Fusion
  Energy}},\ \href
  {http://proquest.umi.com/pqdlink?did=2201203941&sid=1&Fmt=2&clientId=53249&RQT=309&VName=PQD}
  {Ph.D. thesis},\ \bibinfo  {school} {University of California, Los Angeles}
  (\bibinfo {year} {2010})\BibitemShut {NoStop}%
\bibitem [{\citenamefont {Winjum}\ \emph {et~al.}(2010)\citenamefont {Winjum},
  \citenamefont {Fahlen}, \citenamefont {Tsung},\ and\ \citenamefont
  {Mori}}]{Winjum:Effects}%
  \BibitemOpen
  \bibfield  {author} {\bibinfo {author} {\bibfnamefont {B.~J.}\ \bibnamefont
  {Winjum}}, \bibinfo {author} {\bibfnamefont {J.~E.}\ \bibnamefont {Fahlen}},
  \bibinfo {author} {\bibfnamefont {F.~S.}\ \bibnamefont {Tsung}}, \ and\
  \bibinfo {author} {\bibfnamefont {W.~B.}\ \bibnamefont {Mori}},\ }\href
  {\doibase 10.1103/PhysRevE.81.045401} {\bibfield  {journal} {\bibinfo
  {journal} {Phys. Rev. E}\ }\textbf {\bibinfo {volume} {81}},\ \bibinfo
  {pages} {045401} (\bibinfo {year} {2010})}\BibitemShut {NoStop}%
\bibitem [{\citenamefont {Brunner}\ and\ \citenamefont
  {Valeo}(2004)}]{brunner:trapped}%
  \BibitemOpen
  \bibfield  {author} {\bibinfo {author} {\bibfnamefont {S.}~\bibnamefont
  {Brunner}}\ and\ \bibinfo {author} {\bibfnamefont {E.~J.}\ \bibnamefont
  {Valeo}},\ }\href {\doibase 10.1103/PhysRevLett.93.145003} {\bibfield
  {journal} {\bibinfo  {journal} {Phys. Rev. Lett.}\ }\textbf {\bibinfo
  {volume} {93}},\ \bibinfo {pages} {145003} (\bibinfo {year}
  {2004})}\BibitemShut {NoStop}%
\bibitem [{\citenamefont {Yin}\ \emph {et~al.}(2012{\natexlab{a}})\citenamefont
  {Yin}, \citenamefont {Albright}, \citenamefont {Rose}, \citenamefont
  {Bowers}, \citenamefont {Bergen},\ and\ \citenamefont
  {Kirkwood}}]{Yin:Self-Organized}%
  \BibitemOpen
  \bibfield  {author} {\bibinfo {author} {\bibfnamefont {L.}~\bibnamefont
  {Yin}}, \bibinfo {author} {\bibfnamefont {B.~J.}\ \bibnamefont {Albright}},
  \bibinfo {author} {\bibfnamefont {H.~A.}\ \bibnamefont {Rose}}, \bibinfo
  {author} {\bibfnamefont {K.~J.}\ \bibnamefont {Bowers}}, \bibinfo {author}
  {\bibfnamefont {B.}~\bibnamefont {Bergen}}, \ and\ \bibinfo {author}
  {\bibfnamefont {R.~K.}\ \bibnamefont {Kirkwood}},\ }\href {\doibase
  10.1103/PhysRevLett.108.245004} {\bibfield  {journal} {\bibinfo  {journal}
  {Phys. Rev. Lett.}\ }\textbf {\bibinfo {volume} {108}},\ \bibinfo {pages}
  {245004} (\bibinfo {year} {2012}{\natexlab{a}})}\BibitemShut {NoStop}%
\bibitem [{\citenamefont {Fonseca}\ \emph {et~al.}(2002)\citenamefont
  {Fonseca}, \citenamefont {Silva}, \citenamefont {Tsung}, \citenamefont
  {Decyk}, \citenamefont {Lu}, \citenamefont {Ren}, \citenamefont {Mori},
  \citenamefont {Deng}, \citenamefont {Lee}, \citenamefont {Katsouleas},\ and\
  \citenamefont {Adam}}]{fonseca:osiris}%
  \BibitemOpen
  \bibfield  {author} {\bibinfo {author} {\bibfnamefont {R.~A.}\ \bibnamefont
  {Fonseca}}, \bibinfo {author} {\bibfnamefont {L.~O.}\ \bibnamefont {Silva}},
  \bibinfo {author} {\bibfnamefont {F.~S.}\ \bibnamefont {Tsung}}, \bibinfo
  {author} {\bibfnamefont {V.~K.}\ \bibnamefont {Decyk}}, \bibinfo {author}
  {\bibfnamefont {W.}~\bibnamefont {Lu}}, \bibinfo {author} {\bibfnamefont
  {C.}~\bibnamefont {Ren}}, \bibinfo {author} {\bibfnamefont {W.~B.}\
  \bibnamefont {Mori}}, \bibinfo {author} {\bibfnamefont {S.}~\bibnamefont
  {Deng}}, \bibinfo {author} {\bibfnamefont {S.}~\bibnamefont {Lee}}, \bibinfo
  {author} {\bibfnamefont {T.}~\bibnamefont {Katsouleas}}, \ and\ \bibinfo
  {author} {\bibfnamefont {J.}~\bibnamefont {Adam}},\ }in\ \href {\doibase
  10.1007/3-540-47789-6_36} {\emph {\bibinfo {booktitle} {Computational Science
  - ICCS 2002}}},\ \bibinfo {series} {Lecture Notes in Computer Science}, Vol.\
  \bibinfo {volume} {2331},\ \bibinfo {editor} {edited by\ \bibinfo {editor}
  {\bibfnamefont {P.}~\bibnamefont {Sloot}}, \bibinfo {editor} {\bibfnamefont
  {A.}~\bibnamefont {Hoekstra}}, \bibinfo {editor} {\bibfnamefont
  {C.}~\bibnamefont {Tan}}, \ and\ \bibinfo {editor} {\bibfnamefont
  {J.}~\bibnamefont {Dongarra}}}\ (\bibinfo  {publisher} {Springer Berlin /
  Heidelberg},\ \bibinfo {year} {2002})\ pp.\ \bibinfo {pages} {342--351},\
  \bibinfo {note} {{International Conference on Computational Science,
  Amsterdam, Netherlands, April 21-24, 2002}}\BibitemShut {NoStop}%
\bibitem [{\citenamefont {Dawson}(1983)}]{dawson:particle}%
  \BibitemOpen
  \bibfield  {author} {\bibinfo {author} {\bibfnamefont {J.~M.}\ \bibnamefont
  {Dawson}},\ }\href {\doibase 10.1103/RevModPhys.55.403} {\bibfield  {journal}
  {\bibinfo  {journal} {Rev. Mod. Phys.}\ }\textbf {\bibinfo {volume} {55}},\
  \bibinfo {pages} {403} (\bibinfo {year} {1983})}\BibitemShut {NoStop}%
\bibitem [{\citenamefont {Birdsall}\ and\ \citenamefont
  {Langdon}(1985)}]{birdsall:plasma}%
  \BibitemOpen
  \bibfield  {author} {\bibinfo {author} {\bibfnamefont {C.~K.}\ \bibnamefont
  {Birdsall}}\ and\ \bibinfo {author} {\bibfnamefont {A.~B.}\ \bibnamefont
  {Langdon}},\ }\href {http://books.google.com/books?id=7TMbAQAAIAAJ} {\emph
  {\bibinfo {title} {{Plasma Physics Via Computer Simulation}}}},\ The Adam
  Hilger series on plasma physics\ (\bibinfo  {publisher} {McGraw-Hill},\
  \bibinfo {year} {1985})\BibitemShut {NoStop}%
\bibitem [{\citenamefont {Moses}\ \emph {et~al.}(2009)\citenamefont {Moses},
  \citenamefont {Boyd}, \citenamefont {Remington}, \citenamefont {Keane},\ and\
  \citenamefont {Al-Ayat}}]{moses:nif}%
  \BibitemOpen
  \bibfield  {author} {\bibinfo {author} {\bibfnamefont {E.~I.}\ \bibnamefont
  {Moses}}, \bibinfo {author} {\bibfnamefont {R.~N.}\ \bibnamefont {Boyd}},
  \bibinfo {author} {\bibfnamefont {B.~A.}\ \bibnamefont {Remington}}, \bibinfo
  {author} {\bibfnamefont {C.~J.}\ \bibnamefont {Keane}}, \ and\ \bibinfo
  {author} {\bibfnamefont {R.}~\bibnamefont {Al-Ayat}},\ }\href {\doibase
  10.1063/1.3116505} {\bibfield  {journal} {\bibinfo  {journal} {Physics of
  Plasmas}\ }\textbf {\bibinfo {volume} {16}},\ \bibinfo {eid} {041006}
  (\bibinfo {year} {2009})}\BibitemShut {NoStop}%
\bibitem [{\citenamefont {Berenger}(1994)}]{berenger:perfectly}%
  \BibitemOpen
  \bibfield  {author} {\bibinfo {author} {\bibfnamefont {J.-P.}\ \bibnamefont
  {Berenger}},\ }\href {\doibase 10.1006/jcph.1994.1159} {\bibfield  {journal}
  {\bibinfo  {journal} {Journal of Computational Physics}\ }\textbf {\bibinfo
  {volume} {114}},\ \bibinfo {pages} {185 } (\bibinfo {year}
  {1994})}\BibitemShut {NoStop}%
\bibitem [{\citenamefont {Strozzi}\ \emph {et~al.}(2008)\citenamefont
  {Strozzi}, \citenamefont {Williams}, \citenamefont {Hinkel}, \citenamefont
  {Froula}, \citenamefont {London},\ and\ \citenamefont
  {Callahan}}]{Strozzi:Ray}%
  \BibitemOpen
  \bibfield  {author} {\bibinfo {author} {\bibfnamefont {D.~J.}\ \bibnamefont
  {Strozzi}}, \bibinfo {author} {\bibfnamefont {E.~A.}\ \bibnamefont
  {Williams}}, \bibinfo {author} {\bibfnamefont {D.~E.}\ \bibnamefont
  {Hinkel}}, \bibinfo {author} {\bibfnamefont {D.~H.}\ \bibnamefont {Froula}},
  \bibinfo {author} {\bibfnamefont {R.~A.}\ \bibnamefont {London}}, \ and\
  \bibinfo {author} {\bibfnamefont {D.~A.}\ \bibnamefont {Callahan}},\ }\href
  {\doibase 10.1063/1.2992522} {\bibfield  {journal} {\bibinfo  {journal}
  {Physics of Plasmas}\ }\textbf {\bibinfo {volume} {15}},\ \bibinfo {eid}
  {102703} (\bibinfo {year} {2008})},\ \Eprint {http://arxiv.org/abs/0806.0045}
  {arXiv:0806.0045 [physics.plasm-ph]} \BibitemShut {NoStop}%
\bibitem [{\citenamefont {Fried}\ and\ \citenamefont
  {Conte}(1961)}]{fried:plasma}%
  \BibitemOpen
  \bibfield  {author} {\bibinfo {author} {\bibfnamefont {B.}~\bibnamefont
  {Fried}}\ and\ \bibinfo {author} {\bibfnamefont {S.}~\bibnamefont {Conte}},\
  }\href {http://books.google.com/books?id=MclEAAAAIAAJ} {\emph {\bibinfo
  {title} {The plasma dispersion function: the Hilbert transform of the
  Gaussian}}}\ (\bibinfo  {publisher} {Academic Press},\ \bibinfo {year}
  {1961})\BibitemShut {NoStop}%
\bibitem [{\citenamefont {Landau}(1946)}]{Landau:vibrations}%
  \BibitemOpen
  \bibfield  {author} {\bibinfo {author} {\bibfnamefont {L.}~\bibnamefont
  {Landau}},\ }\href@noop {} {\bibfield  {journal} {\bibinfo  {journal}
  {Journal of Physics (USSR)}\ }\textbf {\bibinfo {volume} {10}},\ \bibinfo
  {pages} {25} (\bibinfo {year} {1946})},\ \bibinfo {note} {reproduced in D.
  ter Haar, ed., \textit{Collected papers of L. D. Landau} (Pergamon Press, New
  York, 1965) p. 445-460, and in D. ter Haar, ed., \textit{Men of Physics: L.
  D. Landau}, Vol. 2 (Pergamon Press, New York, 1965).}\BibitemShut {Stop}%
\bibitem [{\citenamefont {Bergman}\ and\ \citenamefont
  {Eliasson}(2001)}]{bergman:linear}%
  \BibitemOpen
  \bibfield  {author} {\bibinfo {author} {\bibfnamefont {J.}~\bibnamefont
  {Bergman}}\ and\ \bibinfo {author} {\bibfnamefont {B.}~\bibnamefont
  {Eliasson}},\ }\href {\doibase 10.1063/1.1358313} {\bibfield  {journal}
  {\bibinfo  {journal} {Physics of Plasmas}\ }\textbf {\bibinfo {volume} {8}},\
  \bibinfo {pages} {1482} (\bibinfo {year} {2001})}\BibitemShut {NoStop}%
\bibitem [{\citenamefont {Bers}, \citenamefont {Shkarofsky},\ and\
  \citenamefont {Shoucri}(2009)}]{bers:relativistic}%
  \BibitemOpen
  \bibfield  {author} {\bibinfo {author} {\bibfnamefont {A.}~\bibnamefont
  {Bers}}, \bibinfo {author} {\bibfnamefont {I.~P.}\ \bibnamefont
  {Shkarofsky}}, \ and\ \bibinfo {author} {\bibfnamefont {M.}~\bibnamefont
  {Shoucri}},\ }\href {\doibase 10.1063/1.3073678} {\bibfield  {journal}
  {\bibinfo  {journal} {Physics of Plasmas}\ }\textbf {\bibinfo {volume}
  {16}},\ \bibinfo {eid} {022104} (\bibinfo {year} {2009})}\BibitemShut
  {NoStop}%
\bibitem [{\citenamefont {Palastro}\ \emph {et~al.}(2010)\citenamefont
  {Palastro}, \citenamefont {Ross}, \citenamefont {Pollock}, \citenamefont
  {Divol}, \citenamefont {Froula},\ and\ \citenamefont
  {Glenzer}}]{palastro:fully}%
  \BibitemOpen
  \bibfield  {author} {\bibinfo {author} {\bibfnamefont {J.~P.}\ \bibnamefont
  {Palastro}}, \bibinfo {author} {\bibfnamefont {J.~S.}\ \bibnamefont {Ross}},
  \bibinfo {author} {\bibfnamefont {B.}~\bibnamefont {Pollock}}, \bibinfo
  {author} {\bibfnamefont {L.}~\bibnamefont {Divol}}, \bibinfo {author}
  {\bibfnamefont {D.~H.}\ \bibnamefont {Froula}}, \ and\ \bibinfo {author}
  {\bibfnamefont {S.~H.}\ \bibnamefont {Glenzer}},\ }\href {\doibase
  10.1103/PhysRevE.81.036411} {\bibfield  {journal} {\bibinfo  {journal} {Phys.
  Rev. E}\ }\textbf {\bibinfo {volume} {81}},\ \bibinfo {pages} {036411}
  (\bibinfo {year} {2010})}\BibitemShut {NoStop}%
\bibitem [{\citenamefont {Tzeng}\ and\ \citenamefont
  {Mori}(1998)}]{tzeng:suppression}%
  \BibitemOpen
  \bibfield  {author} {\bibinfo {author} {\bibfnamefont {K.-C.}\ \bibnamefont
  {Tzeng}}\ and\ \bibinfo {author} {\bibfnamefont {W.~B.}\ \bibnamefont
  {Mori}},\ }\href {\doibase 10.1103/PhysRevLett.81.104} {\bibfield  {journal}
  {\bibinfo  {journal} {Phys. Rev. Lett.}\ }\textbf {\bibinfo {volume} {81}},\
  \bibinfo {pages} {104} (\bibinfo {year} {1998})}\BibitemShut {NoStop}%
\bibitem [{\citenamefont {Decyk}(1987)}]{decyk:simulation}%
  \BibitemOpen
  \bibfield  {author} {\bibinfo {author} {\bibfnamefont {V.~K.}\ \bibnamefont
  {Decyk}},\ }in\ \href@noop {} {\emph {\bibinfo {booktitle} {{1987
  International Conference on Plasma Physics. Joint Conference of the Seventh
  Kiev International Conference on Plasma Theory and the Seventh International
  Congress on Waves and Instabilities in Plasmas. Proceedings of the Invited
  Papers}}}},\ Vol.~\bibinfo {volume} {{2}},\ \bibinfo {editor} {edited by\
  \bibinfo {editor} {\bibfnamefont {A.~G.}\ \bibnamefont {Sitenko}}}\ (\bibinfo
   {publisher} {{World Scientific}},\ \bibinfo {address} {{Singapore,
  Singapore}},\ \bibinfo {year} {{1987}})\ pp.\ \bibinfo {pages}
  {{1075--97}}\BibitemShut {NoStop}%
\bibitem [{\citenamefont {Berger}\ \emph {et~al.}(1998)\citenamefont {Berger},
  \citenamefont {Still}, \citenamefont {Williams},\ and\ \citenamefont
  {Langdon}}]{berger:dominant}%
  \BibitemOpen
  \bibfield  {author} {\bibinfo {author} {\bibfnamefont {R.~L.}\ \bibnamefont
  {Berger}}, \bibinfo {author} {\bibfnamefont {C.~H.}\ \bibnamefont {Still}},
  \bibinfo {author} {\bibfnamefont {E.~A.}\ \bibnamefont {Williams}}, \ and\
  \bibinfo {author} {\bibfnamefont {A.~B.}\ \bibnamefont {Langdon}},\ }\href
  {\doibase 10.1063/1.873171} {\bibfield  {journal} {\bibinfo  {journal}
  {Physics of Plasmas}\ }\textbf {\bibinfo {volume} {5}},\ \bibinfo {pages}
  {4337} (\bibinfo {year} {1998})}\BibitemShut {NoStop}%
\bibitem [{\citenamefont {Strozzi}(2005)}]{strozzi:dissertation}%
  \BibitemOpen
  \bibfield  {author} {\bibinfo {author} {\bibfnamefont {D.~J.}\ \bibnamefont
  {Strozzi}},\ }\emph {\bibinfo {title} {Vlasov Simulations of Kinetic
  Enhancement of Raman Backscatter in Laser Fusion Plasmas}},\ \href
  {http://hdl.handle.net/1721.1/34974} {Ph.D. thesis},\ \bibinfo  {school}
  {Massachusetts Institute of Technology} (\bibinfo {year} {2005})\BibitemShut
  {NoStop}%
\bibitem [{\citenamefont {Wang}\ \emph {et~al.}(2009)\citenamefont {Wang},
  \citenamefont {Michta}, \citenamefont {Lindberg}, \citenamefont {Charman},
  \citenamefont {Martins},\ and\ \citenamefont {Wurtele}}]{wang:feasibility}%
  \BibitemOpen
  \bibfield  {author} {\bibinfo {author} {\bibfnamefont {T.-L.}\ \bibnamefont
  {Wang}}, \bibinfo {author} {\bibfnamefont {D.}~\bibnamefont {Michta}},
  \bibinfo {author} {\bibfnamefont {R.~R.}\ \bibnamefont {Lindberg}}, \bibinfo
  {author} {\bibfnamefont {A.~E.}\ \bibnamefont {Charman}}, \bibinfo {author}
  {\bibfnamefont {S.~F.}\ \bibnamefont {Martins}}, \ and\ \bibinfo {author}
  {\bibfnamefont {J.~S.}\ \bibnamefont {Wurtele}},\ }\href {\doibase
  10.1063/1.3280012} {\bibfield  {journal} {\bibinfo  {journal} {Physics of
  Plasmas}\ }\textbf {\bibinfo {volume} {16}},\ \bibinfo {eid} {123110}
  (\bibinfo {year} {2009})}\BibitemShut {NoStop}%
\bibitem [{\citenamefont {Benisti}, \citenamefont {Morice},\ and\ \citenamefont
  {Gremillet}(2012)}]{benisti:various}%
  \BibitemOpen
  \bibfield  {author} {\bibinfo {author} {\bibfnamefont {D.}~\bibnamefont
  {Benisti}}, \bibinfo {author} {\bibfnamefont {O.}~\bibnamefont {Morice}}, \
  and\ \bibinfo {author} {\bibfnamefont {L.}~\bibnamefont {Gremillet}},\ }\href
  {\doibase 10.1063/1.4729664} {\bibfield  {journal} {\bibinfo  {journal}
  {Physics of Plasmas}\ }\textbf {\bibinfo {volume} {19}},\ \bibinfo {eid}
  {063110} (\bibinfo {year} {2012})},\ \Eprint {http://arxiv.org/abs/1111.1391}
  {arXiv:1111.1391 [physics.plasm-ph]} \BibitemShut {NoStop}%
\bibitem [{\citenamefont {Dewald}\ \emph {et~al.}(2010)\citenamefont {Dewald},
  \citenamefont {Thomas}, \citenamefont {Hunter}, \citenamefont {Divol},
  \citenamefont {Meezan}, \citenamefont {Glenzer}, \citenamefont {Suter},
  \citenamefont {Bond}, \citenamefont {Kline}, \citenamefont {Celeste},
  \citenamefont {Bradley}, \citenamefont {Bell}, \citenamefont {Kauffman},
  \citenamefont {Kilkenny},\ and\ \citenamefont {Landen}}]{dewald:hot}%
  \BibitemOpen
  \bibfield  {author} {\bibinfo {author} {\bibfnamefont {E.~L.}\ \bibnamefont
  {Dewald}}, \bibinfo {author} {\bibfnamefont {C.}~\bibnamefont {Thomas}},
  \bibinfo {author} {\bibfnamefont {S.}~\bibnamefont {Hunter}}, \bibinfo
  {author} {\bibfnamefont {L.}~\bibnamefont {Divol}}, \bibinfo {author}
  {\bibfnamefont {N.}~\bibnamefont {Meezan}}, \bibinfo {author} {\bibfnamefont
  {S.~H.}\ \bibnamefont {Glenzer}}, \bibinfo {author} {\bibfnamefont {L.~J.}\
  \bibnamefont {Suter}}, \bibinfo {author} {\bibfnamefont {E.}~\bibnamefont
  {Bond}}, \bibinfo {author} {\bibfnamefont {J.~L.}\ \bibnamefont {Kline}},
  \bibinfo {author} {\bibfnamefont {J.}~\bibnamefont {Celeste}}, \bibinfo
  {author} {\bibfnamefont {D.}~\bibnamefont {Bradley}}, \bibinfo {author}
  {\bibfnamefont {P.}~\bibnamefont {Bell}}, \bibinfo {author} {\bibfnamefont
  {R.~L.}\ \bibnamefont {Kauffman}}, \bibinfo {author} {\bibfnamefont
  {J.}~\bibnamefont {Kilkenny}}, \ and\ \bibinfo {author} {\bibfnamefont
  {O.~L.}\ \bibnamefont {Landen}},\ }\href {\doibase 10.1063/1.3478683}
  {\bibfield  {journal} {\bibinfo  {journal} {Review of Scientific
  Instruments}\ }\textbf {\bibinfo {volume} {81}},\ \bibinfo {eid} {10D938}
  (\bibinfo {year} {2010})}\BibitemShut {NoStop}%
\bibitem [{\citenamefont {Winjum}\ \emph {et~al.}()\citenamefont {Winjum},
  \citenamefont {Fahlen}, \citenamefont {Tsung},\ and\ \citenamefont
  {Mori}}]{winjum:anomalously}%
  \BibitemOpen
  \bibfield  {author} {\bibinfo {author} {\bibfnamefont {B.~J.}\ \bibnamefont
  {Winjum}}, \bibinfo {author} {\bibfnamefont {J.~E.}\ \bibnamefont {Fahlen}},
  \bibinfo {author} {\bibfnamefont {F.~S.}\ \bibnamefont {Tsung}}, \ and\
  \bibinfo {author} {\bibfnamefont {W.~B.}\ \bibnamefont {Mori}},\ }\href@noop
  {} {\ }\bibinfo {note} {``Anomalously hot electrons due to rescattering of
  stimulated Raman scattering in the kinetic regime," submitted to Phys. Rev.
  Lett.}\BibitemShut {Stop}%
\bibitem [{\citenamefont {Yin}\ \emph {et~al.}(2012{\natexlab{b}})\citenamefont
  {Yin}, \citenamefont {Albright}, \citenamefont {Rose}, \citenamefont
  {Bowers}, \citenamefont {Bergen}, \citenamefont {Kirkwood}, \citenamefont
  {Hinkel}, \citenamefont {Langdon}, \citenamefont {Michel}, \citenamefont
  {Montgomery},\ and\ \citenamefont {Kline}}]{yin:trapping}%
  \BibitemOpen
  \bibfield  {author} {\bibinfo {author} {\bibfnamefont {L.}~\bibnamefont
  {Yin}}, \bibinfo {author} {\bibfnamefont {B.~J.}\ \bibnamefont {Albright}},
  \bibinfo {author} {\bibfnamefont {H.~A.}\ \bibnamefont {Rose}}, \bibinfo
  {author} {\bibfnamefont {K.~J.}\ \bibnamefont {Bowers}}, \bibinfo {author}
  {\bibfnamefont {B.}~\bibnamefont {Bergen}}, \bibinfo {author} {\bibfnamefont
  {R.~K.}\ \bibnamefont {Kirkwood}}, \bibinfo {author} {\bibfnamefont {D.~E.}\
  \bibnamefont {Hinkel}}, \bibinfo {author} {\bibfnamefont {A.~B.}\
  \bibnamefont {Langdon}}, \bibinfo {author} {\bibfnamefont {P.}~\bibnamefont
  {Michel}}, \bibinfo {author} {\bibfnamefont {D.~S.}\ \bibnamefont
  {Montgomery}}, \ and\ \bibinfo {author} {\bibfnamefont {J.~L.}\ \bibnamefont
  {Kline}},\ }\href {\doibase 10.1063/1.3694673} {\bibfield  {journal}
  {\bibinfo  {journal} {Physics of Plasmas}\ }\textbf {\bibinfo {volume}
  {19}},\ \bibinfo {eid} {056304} (\bibinfo {year}
  {2012}{\natexlab{b}})}\BibitemShut {NoStop}%
\bibitem [{\citenamefont {Mallat}(1999)}]{mallat:wavelet}%
  \BibitemOpen
  \bibfield  {author} {\bibinfo {author} {\bibfnamefont {S.}~\bibnamefont
  {Mallat}},\ }\href {http://books.google.com/books?id=hbVOfWQNtB8C} {\emph
  {\bibinfo {title} {A wavelet tour of signal processing}}},\ Wavelet Analysis
  and Its Applications Series\ (\bibinfo  {publisher} {Academic Press},\
  \bibinfo {year} {1999})\BibitemShut {NoStop}%
\bibitem [{\citenamefont {Fahlen}(2010)}]{fahlen:dissertation}%
  \BibitemOpen
  \bibfield  {author} {\bibinfo {author} {\bibfnamefont {J.~E.}\ \bibnamefont
  {Fahlen}},\ }\emph {\bibinfo {title} {Nonlinear Phenomena of Plasma Waves in
  a Kinetic Regime: Frequency Shifts, Packets, and Transverse Localization}},\
  \href
  {http://proquest.umi.com/pqdlink?Ver=1&Exp=08-03-2017&FMT=7&DID=2200212951&RQT=309&attempt=1}
  {Ph.D. thesis},\ \bibinfo  {school} {University of California, Los Angeles}
  (\bibinfo {year} {2010})\BibitemShut {NoStop}%
\bibitem [{\citenamefont {Rose}\ and\ \citenamefont
  {Russell}(2001)}]{rose:self-consistent}%
  \BibitemOpen
  \bibfield  {author} {\bibinfo {author} {\bibfnamefont {H.~A.}\ \bibnamefont
  {Rose}}\ and\ \bibinfo {author} {\bibfnamefont {D.~A.}\ \bibnamefont
  {Russell}},\ }\href {\doibase 10.1063/1.1410111} {\bibfield  {journal}
  {\bibinfo  {journal} {Physics of Plasmas}\ }\textbf {\bibinfo {volume} {8}},\
  \bibinfo {pages} {4784} (\bibinfo {year} {2001})}\BibitemShut {NoStop}%
\bibitem [{\citenamefont {{Strozzi}}\ \emph {et~al.}()\citenamefont
  {{Strozzi}}, \citenamefont {{Williams}}, \citenamefont {{Rose}},
  \citenamefont {{Hinkel}}, \citenamefont {{Langdon}},\ and\ \citenamefont
  {{Banks}}}]{strozzi:characterizing}%
  \BibitemOpen
  \bibfield  {author} {\bibinfo {author} {\bibfnamefont {D.~J.}\ \bibnamefont
  {{Strozzi}}}, \bibinfo {author} {\bibfnamefont {E.~A.}\ \bibnamefont
  {{Williams}}}, \bibinfo {author} {\bibfnamefont {H.~A.}\ \bibnamefont
  {{Rose}}}, \bibinfo {author} {\bibfnamefont {D.~E.}\ \bibnamefont
  {{Hinkel}}}, \bibinfo {author} {\bibfnamefont {A.~B.}\ \bibnamefont
  {{Langdon}}}, \ and\ \bibinfo {author} {\bibfnamefont {J.~W.}\ \bibnamefont
  {{Banks}}},\ }\href {\doibase 10.1063/1.4767644} {\ 10.1063/1.4767644},\
  \bibinfo {note} {``Characterizing Electron Trapping Nonlinearity in Langmuir
  Waves," submitted to Phys. Plasmas},\ \Eprint
  {http://arxiv.org/abs/1208.3864} {arXiv:1208.3864 [physics.plasm-ph]}
  \BibitemShut {NoStop}%
\bibitem [{\citenamefont {{B. J. Albright}}\ \emph {et~al.}(2006)\citenamefont
  {{B. J. Albright}}, \citenamefont {{W. Daughton}}, \citenamefont {{L. Yin}},
  \citenamefont {{K.J. Bowers}}, \citenamefont {{J.L. Kline}}, \citenamefont
  {{D.S. Montgomery}},\ and\ \citenamefont {{J.C.
  Fern\'andez}}}]{albright:particle-in-cell}%
  \BibitemOpen
  \bibfield  {author} {\bibinfo {author} {\bibnamefont {{B. J. Albright}}},
  \bibinfo {author} {\bibnamefont {{W. Daughton}}}, \bibinfo {author}
  {\bibnamefont {{L. Yin}}}, \bibinfo {author} {\bibnamefont {{K.J. Bowers}}},
  \bibinfo {author} {\bibnamefont {{J.L. Kline}}}, \bibinfo {author}
  {\bibnamefont {{D.S. Montgomery}}}, \ and\ \bibinfo {author} {\bibnamefont
  {{J.C. Fern\'andez}}},\ }\href {\doibase 10.1051/jp4:2006133051} {\bibfield
  {journal} {\bibinfo  {journal} {J. Phys. IV France}\ }\textbf {\bibinfo
  {volume} {133}},\ \bibinfo {pages} {253} (\bibinfo {year}
  {2006})}\BibitemShut {NoStop}%
\bibitem [{\citenamefont {B\'enisti}\ \emph {et~al.}(2009)\citenamefont
  {B\'enisti}, \citenamefont {Strozzi}, \citenamefont {Gremillet},\ and\
  \citenamefont {Morice}}]{benisti:nonlinear}%
  \BibitemOpen
  \bibfield  {author} {\bibinfo {author} {\bibfnamefont {D.}~\bibnamefont
  {B\'enisti}}, \bibinfo {author} {\bibfnamefont {D.~J.}\ \bibnamefont
  {Strozzi}}, \bibinfo {author} {\bibfnamefont {L.}~\bibnamefont {Gremillet}},
  \ and\ \bibinfo {author} {\bibfnamefont {O.}~\bibnamefont {Morice}},\ }\href
  {\doibase 10.1103/PhysRevLett.103.155002} {\bibfield  {journal} {\bibinfo
  {journal} {Phys. Rev. Lett.}\ }\textbf {\bibinfo {volume} {103}},\ \bibinfo
  {pages} {155002} (\bibinfo {year} {2009})}\BibitemShut {NoStop}%
\bibitem [{\citenamefont {Yampolsky}\ and\ \citenamefont
  {Fisch}(2009)}]{Yampolsky:Simplified}%
  \BibitemOpen
  \bibfield  {author} {\bibinfo {author} {\bibfnamefont {N.~A.}\ \bibnamefont
  {Yampolsky}}\ and\ \bibinfo {author} {\bibfnamefont {N.~J.}\ \bibnamefont
  {Fisch}},\ }\href {\doibase 10.1063/1.3160604} {\bibfield  {journal}
  {\bibinfo  {journal} {Physics of Plasmas}\ }\textbf {\bibinfo {volume}
  {16}},\ \bibinfo {eid} {072104} (\bibinfo {year} {2009})}\BibitemShut
  {NoStop}%
\bibitem [{\citenamefont {Lindberg}, \citenamefont {Charman},\ and\
  \citenamefont {Wurtele}(2007)}]{Lindberg:Self-consistent}%
  \BibitemOpen
  \bibfield  {author} {\bibinfo {author} {\bibfnamefont {R.~R.}\ \bibnamefont
  {Lindberg}}, \bibinfo {author} {\bibfnamefont {A.~E.}\ \bibnamefont
  {Charman}}, \ and\ \bibinfo {author} {\bibfnamefont {J.~S.}\ \bibnamefont
  {Wurtele}},\ }\href {\doibase 10.1063/1.2801714} {\bibfield  {journal}
  {\bibinfo  {journal} {Physics of Plasmas}\ }\textbf {\bibinfo {volume}
  {14}},\ \bibinfo {eid} {122103} (\bibinfo {year} {2007})},\ \Eprint
  {http://arxiv.org/abs/physics/0701288} {arXiv:physics/0701288
  [physics.plasm-ph]} \BibitemShut {NoStop}%
\bibitem [{\citenamefont {Fahlen}\ \emph {et~al.}(2011)\citenamefont {Fahlen},
  \citenamefont {Winjum}, \citenamefont {Grismayer},\ and\ \citenamefont
  {Mori}}]{fahlen:transverse}%
  \BibitemOpen
  \bibfield  {author} {\bibinfo {author} {\bibfnamefont {J.~E.}\ \bibnamefont
  {Fahlen}}, \bibinfo {author} {\bibfnamefont {B.~J.}\ \bibnamefont {Winjum}},
  \bibinfo {author} {\bibfnamefont {T.}~\bibnamefont {Grismayer}}, \ and\
  \bibinfo {author} {\bibfnamefont {W.~B.}\ \bibnamefont {Mori}},\ }\href
  {\doibase 10.1103/PhysRevE.83.045401} {\bibfield  {journal} {\bibinfo
  {journal} {Phys. Rev. E}\ }\textbf {\bibinfo {volume} {83}},\ \bibinfo
  {pages} {045401} (\bibinfo {year} {2011})}\BibitemShut {NoStop}%
\end{thebibliography}%

\end{document}